\documentclass{aa}   

\usepackage{graphicx}
\usepackage{url}
\usepackage{color}
\bibpunct{(}{)}{;}{a}{}{,} 
\usepackage{psfrag}
\usepackage{version}
\usepackage{txfonts}
\usepackage{float}
\usepackage[colorlinks=true,linkcolor=blue,citecolor=blue,urlcolor=black]{hyperref}
\usepackage[normalem]{ulem}

\bibpunct{(}{)}{;}{a}{}{,} 

\begin{document} 

	\title{Nano carbon dust emission in proto-planetary disks: 
    the aliphatic-aromatic components}

	\author{T. Boutéraon \inst{1},
          E. Habart\inst{1},
          N. Ysard \inst{1},
          A.P. Jones \inst{1},
          E. Dartois \inst{2},
          T. Pino \inst{2}
          }

	\institute{\inst{1}Institut d’Astrophysique Spatiale (IAS), UMR 8617-CNRS Université Paris Sud, 91405 Orsay, France \\             
               \inst{2}Institut des Sciences Moléculaires d’Orsay (ISMO), CNRS, Univ Paris Sud, Université Paris-Saclay, 91405 Orsay, France \\
               \email{thomas.bouteraon@ias.u-psud.fr}
               }

	\date{Preprint online version: January 22, 2019}

 
  \abstract
   {In the interstellar medium, carbon (nano-)grains are a major component of interstellar dust. This solid phase is more vulnerable to processing and destruction than its silicate counterpart. It exhibits a complex, size-dependent evolution due to interactions within different radiative and dynamical environments. Infrared signatures of these nanocarbon grains are seen in a large number of disks around Herbig HAeBe stars.}
   {We probe the composition and evolution of carbon nano-grains at the surface of (pre-)transitional protoplanetary disks around Herbig stars.}
   {We present spatially resolved infrared emission spectra obtained with NAOS CONICA at the VLT in the 3-4 $\mu$m range with a spatial resolution of 0.1", which allow us to trace aromatic, olefinic andhttps://www.overleaf.com/project/5c46c1dbf80925580ab5dd2d aliphatic bands which are attributed to sub-nanometer hydrocarbon grains. We apply a gaussian fitting to analyse the observed spectral signatures. Finally, we propose an interpretation in the framework of the The Heterogeneous dust Evolution Model of Interstellar Solids (THEMIS).}
   {We show the presence of several spatially extended spectral features, related to aromatic and aliphatic hydrocarbon material in disks around Herbig stars, from $\sim$ 10 to 50-100 au, and even in inner gaps devoided of large grains. The correlation and constant intensity ratios between aliphatic and aromatic CH stretching bands suggest a common nature of the carriers. Given their expected high destruction rates due to UV photons, our observations suggest that they are continuously replenished at the disk surfaces.
   }
   {}

\titlerunning{Carbonaceous dust grains emission}
\maketitle

%

\section{Introduction \label{sec_intro}}

Carbonaceous dust is a major component of the solid matter present in the interstellar medium (ISM). It is observed in a wide variety of environments, both in emission and extinction. Much more sensitive to energetic events than its silicate counterpart, carbonaceous dust has been shown to be highly processed in UV irradiated regions and in shocks \citep[e.g.][]{jones_carbon_1990,dartois_diffuse_2004,dartois_organic_2004,pino_6.2_2008,mennella_activation_2008,godard_ion_2011,micelotta_polycyclic_2010,micelotta_polycyclic_2010-1}. It is thus expected to be sensitive to the local physical conditions in the environments where it is observed. 
Processing of carbonaceous dust is expected to be reflected in the aromatic and aliphatic near- to mid-IR emission bands that are attributed to sub-nanometers polycyclic aromatic hydrocarbon-like molecules (PAHs) \citep{leger_identification_1984} or hydrogenated amorphous carbons (a-C:H) \citep[and references therein]{jones_carbon_1990}. 
a-C(:H), including both H-poor a-C and H-rich a-c:H, materials are a broad family of compounds including various proportion of polyaromatic units, of various sizes, linked by olefinic and aliphatic bridges. This component, when observed in emission, associated with the aromatic phase, should thus exhibit a series of bands between 3.4 and 3.6$~\mu$m, in addition to the aromatic 3.3~$\mu$m band \citep[e.g.,][]{geballe_spectroscopy_1985, jourdain_de_muizon_three_1990, jourdain_de_muizon_polycyclic_1990, joblin_spatial_1996, goto_spatially_2003, pilleri_mixed_2015}, as well as, bands at 6.9 and 7.3$~\mu$m \citep{pino_6.2_2008,carpentier_nanostructuration_2012}.

Aromatic Infrared Bands (AIBs) have been observed using ISO, Spitzer, AKARI and ground-based observations towards proto-planetary disks around pre-main-sequence stars \citep[e.g.,][]{brooke_dust_1993,meeus_iso_2001,acke_iso_2004,habart_pahs_2004,sloan_mid-infrared_2005,geers_c2d_2006,geers_spatial_2007,geers_spatially_2007,keller_pah_2008,acke_spitzers_2010,maaskant_polycyclic_2014,seok_polycyclic_2017}. 
Based on Spitzer data, the detection rate is of $\sim$70\% in Herbig Ae stars, $\sim$50\% in Herbig Be stars \citep{acke_spitzers_2010}, and $\sim$10\% in T Tauri stars \citep{furlan_survey_2006,geers_c2d_2006}. AIBs are detected only towards a few debris disks \citep{chen_spitzer_2006}. Furthermore, mid-IR bands attributed to C-H aliphatic bonds at 6.9 and 7.3~$\mu$m have been observed in $\sim$55\% of Herbig stars \citep{acke_spitzers_2010}. 
The study of the aromatic/aliphatic band ratio in the mid-IR Spitzer spectra \citep{acke_spitzers_2010} suggests that strong UV fluxes reduce the aliphatic component and magnify the spectral signature of the aromatic molecules in the IR spectra. However, dehydrogenation caused by UV photons \citep{munoz_caro_uv_2001} also reduces C-H aromatic signatures.
Finally, fullerenes have been detected in one disk \citep{roberts_detection_2012}. 

Bands between 3.4 and 3.6$~\mu$m have also been detected in some Herbig Ae/Be (HAeBe) stars \citep[e.g.,][]{acke_resolving_2006,habart_spatially_2006}. Few of them (2 to 3) show peculiar strong features that peak at 3.43 and 3.53$~\mu$m \citep[e.g.,][]{blades_observations_1980,guillois_diamond_1999,van_kerckhoven_nanodiamonds_2002}. Using adaptative optics high angular resolution spectroscopic observations of the HAeBe star HD 97048 and Elias 1, the emission in the strong features at 3.43 and 3.53~$\mu$m was spatially resolved \citep{habart_pahs_2004,goto_spatially_2009}. Several studies have proposed attributing these features to surface C-H stretching modes on 10-50 nm or larger diamond particles \citep[e.g,][]{guillois_diamond_1999,sheu_laboratory_2002,jones_surface_2004,pirali_infrared_2007}. 
On the other hand, some disks show relatively weak but distinguishable features between 3.4 and 3.6~$\mu$m at 3.40, 3.46, 3.51, and 3.56~$\mu$m \citep[e.g.][]{sloan_variations_1997}(e.g.  Sloan et al 1997) which may be attributed to aliphatic components. These features were also detected towards Photo-Dissociation Regions (PDRs) in star forming regions, reflection nebulae and (proto)planeatary nebulae \citep[e.g.,][]{geballe_spectroscopy_1985,jourdain_de_muizon_three_1990,jourdain_de_muizon_polycyclic_1990,joblin_spatial_1996,goto_spatially_2003,pilleri_mixed_2015}.
Toward a proto-planetary nebula, \citet{goto_spatially_2003} found that the relative intensity of the aliphatic feature at 3.4~$\mu$m to the aromatic feature at 3.3~$\mu$m decreases with the distance from the star. They suggested that thermal processing is likely to account for the spectral variation, reflecting the history of the planetary nebula.
Towards the NGC 7023 North West PDR, \citet{pilleri_mixed_2015} traced the evolution of the 3.3 and 3.4~$\mu$m bands. They showed that the intensity of the 3.3~$\mu$m band relative to the total PAH emission increases with the UV flux, while the relative contribution of the 3.4~$\mu$m band decreases with the UV flux. The UV photo-processing of Very Small Grains leading to PAHs with attached aliphatic sidegroups in the lower UV flux region, and then becoming PAHs in the higher UV flux region was suggested to explain the spectral variation.
Variations in the ratio of the 3.3 to 3.4 micron bands raise the question of the nature and stability of the carriers of these bands and of their excitation mechanisms as a function of the local physical conditions. 

Studying the aromatic and aliphatic bands in proto-planetary disks, where the carbon nano-grains are likely highly processed, is thus important to understand the mechanisms at work on carbon dust. 
Photo-induced processes may be important at the disk surfaces.
Carbon nano-grains, that escape settling, are very well coupled to the gas in the upper disk layers. The AIBs emission arises from the very thin upper disk surface layers where UV photons from the central star penetrate, which can be interpreted as a PDR. 
Because of the stochastic heating mechanism, the emission in the bands is spatially extended \citep[e.g.]{habart_spatially_2006} and thus makes it possible to probe the composition of the disk surface material over large distances, as well as, the geometry of the disks. 
High AIBs-to-stellar luminosity ratios are observed in targets with a flared disk \citep[e.g.,][]{meeus_iso_2001,habart_pahs_2004,keller_pah_2008,acke_spitzers_2010}. Observations of AIBs provide one of the most striking indications for flared disks around HAeBe stars \citep[e.g.,][]{lagage_anatomy_2006,berne_very_2015}.
Furthermore, AIBs have been detected in the inner disk cavities, where on-going planet formation could have started \citep[e.g.,][]{geers_spatial_2007,maaskant_identifying_2013,maaskant_polycyclic_2014,kraus_resolving_2013,schworer_resolved_2017}. Recently, near-IR interferometric observations of HAeBe stars tracing the inner part of the warm dusty disks revealed unusually extended NIR emission \citep[out to $\sim$10 AU][]{klarmann_interferometric_2017,kluska_multi-instrument_2018}. This suggests the presence of stochastically heated very small (sub-)nanometer grains in the inner disk parts. 
Finally, carbon nano-grains could play, as in PDRs, a key role in gas heating via the photoelectric effect and, thus influence the disk vertical structure \citep[e.g.,][]{meeus_observations_2012}.
Due to their very large effective surface area, nano-grains are also expected to have an important role in the formation of key molecules (e.g., H$_2$, hydrocarbon molecules) and charge balance. Hence, a better knowledge of them is essential to understand the disk physics and chemistry. 	

This study focuses on the near-IR emission between 3 and 4 $\mu$m of spatially resolved spectra around young Herbig stars orbited by a proto-planetary disk and observed with the NAOS CONICA (NACO) instrument at the Very Large Telescope (VLT) in Chile. This is the first spatially resolved study of the weak bands in the 3 to 4 $\mu$m-range in protoplanetary disks. Our aim is to use the aromatic and aliphatic signatures detected at the disk surfaces to characterise the size and composition of the sub-nanometer carbonaceous grains.
In section \ref{sec_versat}, we discuss the diversity of the carbonaceous signatures. 
In section \ref{sec_selobj}, we give the astrophysical parameters and the main characteristics of the observed disks. 
In section \ref{sec_reduc}, we describe the observations and data reduction. 
In section \ref{sec_results}, we present the observational results, the spectral signatures fitting method and the identified bands.
In section \ref{sec_mod}, we briefly describe the THEMIS framework, a global model, in which dust composition, evolution and signatures can be analysed.
In section \ref{sec_conclusion}, we discuss the results and their astrophysical implications. 

\section{The variability of the carbonaceous signatures \label{sec_versat}}

\begin{figure*}
\centering
\includegraphics[width=17cm]{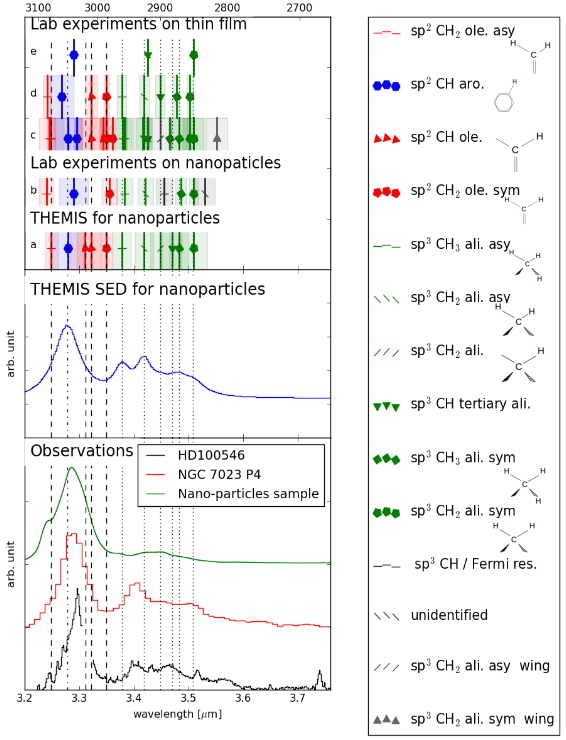}
\caption{Top panel shows the band centre (line) and width (transparent box) of the spectral signatures related to a-C:H materials obtained in laboratory experiments on thin film or on nanoparticles in the mid-IR range where C-H stretching vibrational modes are observed. In blue is shown the aromatic compound, in red olefinic bonds, in green aliphatic ones and in grey other non-assigned signatures. The signatures used in the THEMIS model are based on these experiments and are drawn in row {\it a} and are extended by vertical lines for aromatics (dash-dotted), olefinics (dashed), aliphatics (dotted). Middle panel shows the spectral energy distribution (SED) for an H-poor, aromatic-rich grain population from the THEMIS model and calculated with DustEM. The bottom panel show characteristic spectrum from our study that shows the whole diversity observed in this range around the Herbig star HD 100546. Spectra from the PDR NGC 7023 \citep{pilleri_mixed_2015} and nano-particles \citep{carpentier_nanostructuration_2012} (related to the study of \cite{pino_6.2_2008}) are also plotted for the comparison. In the right panel, the signatures are labelled and a schematic view of vibrational modes is given. {\it a}: \citet{jones_global_2017}, {\it b}: \citet{pino_6.2_2008}, {\it c}: \citet{dartois_diffuse_2004}, {\it d}:  \citet{ristein_comparative_1998}, {\it e}: \citet{dischler_amorphous_1987}.}
\label{fig_versat}
\end{figure*}
   
Fig. \ref{fig_versat} presents the diversity of features related to hydrogenated amorphous carbonaceous (a-C:H) materials in the 3-4~$\mu$m near-infrared (IR) range where C-H stretching vibrational modes are observed. These vibrational modes are particularly interesting since they characterise bonds between carbon and hydrogen atoms varying according to their local environment (closest neighbours).

Fig. \ref{fig_versat} illustrates these variations for the related signatures of two major types of laboratory experiments. Four experiments concern a-C:H deposited on thin film samples  \citep{dartois_ultraviolet_2005,dartois_diffuse_2004,ristein_comparative_1998,dischler_amorphous_1987} and one experiment focuses on soot nano-particles, having disordered carbon nano-structurations, produced in flames \citep{pino_6.2_2008}. Analysis had been made using IR absorption spectroscopy. The experiments of these studies give an overview of the variability of the features related to carbonaceous materials and highlight the difficulty of signature assignments. It should be noticed that the band origin assignment in the 3.22 - 3.32~$\mu$m range is still quite uncertain due to the presence of four or five overlapping  vibrations in this region. 

Table \ref{tab_versat} gives assumed signature assignments, their band centre and width ranges, both in wavelengths and wavenumbers. These modes correspond to C-H stretching (str) vibrational modes.
We distinguish three main kinds of bonds: aromatics, olefinics, and aliphatics. Representation of such vibration modes are given in Fig. \ref{fig_versat}.
The aromatic band at 3.3~$\mu$m corresponds to the vibration of hydrogen bonded to a carbon included in an aromatic cycle and labeled sp$^2$ (hybridization state) CH aromatic (aro.). Its band centre varies from 3.268 to 3.295~$\mu$m and its width from 0.005 to 0.057~$\mu$m.
Olefinic (ole.) modes correspond to the vibration of CH or CH$_2$ group where the carbon atom is engaged in a C=C bond. The CH$_2$ asymmetric (asy) mode varies from 3.240 to 3.249~$\mu$m with a width between 0.014 and 0.045~$\mu$m. It is less variable than its symmetrical (sym) counterpart, for which the band centre varies from 3.350 to 3.396 $\mu$m and the width between 0.003 and 0.028~$\mu$m. 
A third olefinic mode involving only CH group occurs at an intermediate wavelength between 3.311 and 3.344~$\mu$m with width variations between 0.008 and 0.062~$\mu$m. 
Theses modes are related to a sp$^2$ planar hybridization, where one carbon atom forms bonds with three other atoms.

Two types of aliphatic (ali.) modes can be distinguished in the CH stretch region: those originated in the CH$_3$ and CH$_2$ functional group, with their asymmetric and symmetric characters. The CH$_3$ aliph asy mode varies from 3.378 to 3.84~$\mu$m and its width between 0.008 and 0.033~$\mu$m, while the CH$_2$ aliph asy mode varies from 3.413 to 3.425~$\mu$m and its width between 0.006 and 0.034~$\mu$m. At longer wavelength, the CH$_3$ aliph sym mode varies from 3.466 to 3.486~$\mu$m and its width between 0.006 and 0.034~$\mu$m while the CH$_2$ aliph sym mode varies from 3.503 to 3.509~$\mu$m and its width between 0.006 and 0.051~$\mu$m.
A fifth mode related to the aliphatic CH can be found at an intermediate wavelength between symmetric and asymmetric modes. It is labeled as tertiarty and its band centre varies from 3.425  to 3.47~$\mu$m and its width between 0.006 and 0.036~$\mu$m.
All aliphatic modes have a sp$^3$ hybridization in which carbon atoms are bonded to four other atoms.

In the following, observational data will be compared to the THEMIS model \citep{jones_global_2017,jones_evolution_2013}, which takes into account vibrational modes coming from the thin film experiments described above. The bands included in THEMIS (assignment and characteristics) are also plotted in Fig. \ref{fig_versat} for comparison. A typical spectrum from the disk observation and one from the PDR NGC 7023 \citep{pilleri_mixed_2015} are also plotted for comparison. Obvious differences between lab data, observations, and the model are found, which will be discussed in Sect. \ref{sec_mod}.

\begin{table*}
\caption{Band centre ($\lambda_0$, $\nu_0$) and full width at half maximum (FWHM) variations in laboratory experiments.}             
\label{tab_versat}      
\centering        
\begin{tabular}{c c c c c}     
\hline \hline 
 band & $\lambda_{0}$ [$\mu$m]  & FWHM [$\mu$m] & $\nu_0$ [cm$^{-1}$] & FWHM [cm$^{-1}$]\\
\hline
sp$^2$ CH$_2$ ole. asy str & 3.240 - 3.249 & 0.014 - 0.045 & 3089 - 3078 &  13.3 - 42.4\\
 sp$^2$ CH aro & 3.268 - 3.295  & 0.005 - 0.057 & 3060 - 3035 & 5 - 53.1 \\
 sp$^2$ CH ole. & 3.311 - 3.344 &  0.008 - 0.062 & 2990 - 3020  & 7 - 56.3  \\
 sp$^2$ CH$_2$ ole. sym str & 3.350 - 3.396  & 0.003 - 0.028 & 2985 - 2945 & 3 - 25 \\
 sp$^2$ CH$_3$ ali. asy str & 3.378 - 3.384 & 0.008 - 0.033 & 2960 - 2955  & 7 - 29.3 \\
 sp$^3$ CH$_2$ ali. asy str & 3.413 - 3.425 & 0.006 - 0.034 & 2930 -2920  & 5 - 28.9 \\
 sp$^3$ CH tertiarty ali. / Fermi resonance & 3.425 -3.47 & 0.006 - 0.036 & 2920 - 2882 & 5 - 30 \\
 sp$^3$ CH$_3$ ali. sym str & 3.466 - 3.486 & 0.006 - 0.034 & 2885 - 2869 & 5 - 27.8 \\
 sp$^3$ CH$_2$ ali. sym str & 3.503 - 3.509 & 0.006 - 0.051 & 2855 - 2850 & 5 - 41.8 \\
 sp$^3$ CH$_2$ ali. sym str wing & 3.552  & 0.039 - 0.068 & 2815 & 31 - 53.6\\
\hline                  
\end{tabular}
\end{table*}

\section{ Selected (pre-)transitional disks \label{sec_selobj}}

Our sample consists of four well-known Herbig Ae/Be stars, all with evidence of a circumstellar disk thought to be in transition from a gas-rich proto-planetary disk to a gas-depleted debris disk. 
The four disks show clear evidence of flaring from their spectral energy distribution modelling (\citet{meeus_iso_2001}), and IR and millimeter interferometry observations (detailed descriptions of the disks and the associated references are given in Appendix \ref{app_disks}). Furthermore, it has recently become clear that these four disks show many spatial structures, such as annular gaps and rings in the large grain emission distribution near the mid-plane as seen in millimeter and in the near-IR stellar scattered light by $\mu$m grains. Large scale spiral arms are also seen in scattered light. In the innermost few astronomical units (AU), evidence for a radial gap separating an inner disk (from $\sim$0.1 to few AU) and the outer main disk (from $\sim$10 to 100 AU or more) was found using near- and mid-IR observations.  Some of these structures could be tracing on-going planet formation. Planet candidates have been found around two of our disks: HD~100546 \citep[e.g.,][]{quanz_confirmation_2015,currie_resolving_2015} and HD~169142 \citep{biller_enigmatic_2014,reggiani_discovery_2014,osorio_imaging_2014}. However, as discussed by several authors, the observed disk structures could also result from gravitational perturbations by binary stars, disk dynamical evolution (e.g., gas accretion from the surrounding molecular cloud, local maximum pressure) or dust evolution.

These disks have been observed at different wavelengths, using ISO, Spitzer and ground-based spectroscopy, all showing aromatic emission features and  possible aliphatic emission features \citep{meeus_iso_2001,acke_iso_2004,sloan_mid-infrared_2005,habart_spatially_2006,geers_spatially_2007,keller_pah_2008,acke_spitzers_2010,seok_polycyclic_2017}.  
In Table \ref{tab_selobj}, we report (i) the astrophysical parameters of each star (spectral type, effective temperature, luminosity, mass, age, distance, and G$_0$, i.e., far-ultraviolet (FUV) flux strength at 50 AU from the star in terms of the Habing field \citep{habing_interstellar_1968}, and (ii) the dust spectral characteristics (presence of aromatic and aliphatic). We also give the strength of the 3.3~$\mu$m aromatic emission feature measured from the ISO spectra of \citet{meeus_iso_2001,van_kerckhoven_nanodiamonds_2002}. Estimations of the radial spatial extension of the 3.3~$\mu$m aromatic emission feature (full-width at half-maximum, FWHM) measured from ground-based spatially resolved observations are also reported.

\begin{table*}
\caption{Top part: Astrophysical parameters of the selected disks. The standard star used for data reduction is HR6572.
Spectral type, effective temperature, luminosity, mass, age \citep{seok_polycyclic_2017}. Distance \citep{gaia_collaboration_gaia_2018}. $G_0$: far-ultraviolet (FUV) flux strength at 50 AU from the star expressed in units of the average interstellar radiation field, 1.6 $\times 10^{-3}$ erg s$^{-1}$ cm$^{-2}$ \citep{habing_interstellar_1968}. Integrated strength of the aromatic 3.3 $\mu$m feature \citep{habart_pahs_2004}. Estimation of the radial spatial extension of the 3.3~$\mu$m band (full width at half-maximum, FWHM) for HD 100546, HD 100453 and HD 169142 \citep{geers_spatially_2007,habart_spatially_2006}. FWHM for HD 17918 (this work) Bottom part: Summary of the observations. SR, Strehl ratio or coherent energy. $r_0$, Fried parameter. $L_0$, outer scale.}     
\label{tab_selobj}      
\centering        
\begin{tabular}{c c c c c}     
  \hline \hline 
  &   HD 100546 &   HD 100453 &   HD 169142 &   HD 179218 \\
  \hline
  Spec. type  & B9Vne & A9Ve & B9V & A0Ve	  \\
  Temp. [$K$] & 10500 & 7600 & 8250 & 9640 \\
  Lum. [$L_{\odot}$] & 32  & 10 & 8.55 & 182 \\
  M [$M_{\odot}$] & 2.4  & 1.8 & 1.69 & 3.66 \\
  Age [$Myr$] & > 10 & 15 (10 ?) & 6 & 1.08 \\
  d [$pc$]  & 109 & 104 & 117 & 245 \\
  $G_0$ at 50 AU  &  4.2 $\times 10^{6}$ & 2.4 $\times 10^{5}$  &  3.4 $\times 10^{5}$  &  1.6 $\times 10^{7}$\\
  $I_{3.3}$ [$10^{-14}$ $W/m^2$] & 2.5$\pm$0.5 & 1.3$\pm$0.2 &  1$\pm$0.2  & 1.7$\pm$0.2\\
  FWHM [AU]  & 12 (up to 50) & 20 & 4	3 & 35  \\
\hline
  Date & 2005/03/26 & 2005/03/30 & 2005/05/05 & 2005/07/14 \\
  Airmass & 1.56 & 1.17 & 1.01 & 1.32 \\
  SR [$\%$] & 46 & 38 & 37 & 53 \\
  $r_0$ [$cm$] & 16 & 8 & 11 & 14 \\
  $L_0$ [$m$] & 25 & 19 & 18 & 17 \\
  Seeing [$"$] & 0.66 & 1.17 & 0.93 & 0.69 \\
\hline
\end{tabular}

\end{table*}

\section{ Observations and data reduction \label{sec_reduc}}

Observations were performed using a long slit in the L-band, between 3.20 and 3.76~$\mu$m with the adaptive optics system NAOS-CONICA (NACO) at the VLT. The on sky projection of the slit is 28"-long and 0.086"-wide that corresponds to the diffraction limit in this wavelength range. The pixel scale is 0.0547" and the spectral resolution is $R=\lambda$ / $\Delta$ $\lambda$ \textasciitilde 1000. We took several slit positions, one centred on the star and the other slits shifted by a half width. Seven or nine positions allowed to extract spectra on an area star-centred of 2"x0.258" or 2"x0.354", respectively. 
The long slit was aligned in the north-south direction, except for HD 100546. For HD 100546, the long slit was aligned with the major axis of the disk as resolved in scattered light \citep{augereau_hst/nicmos2_2001,grady_disk_2001} with a position angle of $\sim 160$ degrees measured east of north.
The integration time per slit was 3 minutes. 
The data set reference is 075.C-0624(A) and observations characteristics are summarised in Table \ref{tab_selobj}. 

Ground-based near-IR spectroscopic observations involve the use of a standard procedure to subtract the atmospheric contribution. 
For each disk, we employed the standard chopping/nodding technique with a throw of $\sim$9" in the north-south direction in order to correct from the atmospheric and instrumental background. 
Telluric bands due to the atmosphere were used for the wavelength calibration of spectra. They were removed using standard stars observed during each data set acquisition. 
To compensate for the airmass variation between the observations of the disk and the standard star, a scaling factor was applied to the standard star spectrum to optimize telluric band subtraction. 
 

\section{ Results \label{sec_results}}

\subsection{Extraction, decomposition and characteristic of the spectra \label{subsec_dec}}

For the four disks, spectra were extracted from pixels for each slit on both sides of the central star, covering an area of  2"$\times$0.258" or 2"$\times$0.354" star-centred. 
Since we aim to focus on the carbonaceous dust emission evolution we extract the features from the original data by subtracting an underlying continuum.
The continuum is subtracted with a polynomial of order three considering ranges where bands are absent,\textit{ i.e}. 3.6 - 3.7~$\mu$m range, and 3.2~$\mu$m. The intensity at 3.35~$\mu$m was used to constrain the polynomial so that the continuum is not higher than the signal.
The spectra show similar structures for the four disks: the 3.3~$\mu$m aromatic feature, at least five features between 3.4 and 3.6~$\mu$m (cf Fig. \ref{fig_zdec}), and hydrogen recombination lines Pfund 9 and 8 at 3.297 and 3.741~$\mu$m.

To interpret the  shape and the evolution of the features moving away from the central star, we consider a heterogeneous widening assuming the presence of several carriers associated with the same signature and make a Gaussian decomposition of the spectra based on the experiments discussed in Sect. \ref{sec_versat}. The size, temperature, structure, and even composition of the various carriers imply that the signature is composed of multiple components. A good approximation of these signatures is thus a series of gaussian profiles. We use eight Gaussians: 
\begin{itemize}
\item 3.3~$\mu$m : related to the C-H stretching mode of aromatics
\item 3.4, 3.43, 3.46, 3.52, and 3.56~$\mu$m  most probably associated with aliphatic materials.
\item 3.297 and 3.741~$\mu$m  which are the recombination lines Pfund 9 and 8 of hydrogen.
\end{itemize}
The decomposition is made on the continuum subtracted spectra expressed in wavenumber. The initial parameters and their tolerance range are summarised in Table \ref{tab_decparams}. A zoom of the decomposition is presented in Fig. \ref{fig_zdec}.

\begin{table}
\caption{Initial parameters of the gaussian decomposition method.}             
\label{tab_decparams}      
\centering        
\begin{tabular}{c c c}     
  \hline \hline 
  Feature & Band centre [cm$^{-1}$] & Standard deviation [cm$^{-1}$] \\
  \hline
3.3 $\mu$m & 3040 $\pm$ 10 & 25 $\pm$ 10 \\
3.4 $\mu$m & 2941 $\pm$ 10 & 15 $\pm$ 5 \\
3.43 $\mu$m & 2915 $\pm$ 10 & 10 $\pm$ 5 \\
3.46 $\mu$m & 2890 $\pm$ 10 & 15 $\pm$ 5 \\
3.52 $\mu$m & 2841 $\pm$ 10 & 15 $\pm$ 5 \\
3.56 $\mu$m & 2809 $\pm$ 10 & 20 $\pm$ 10 \\
H Pfund 9 & 3027 $\pm$ 5 & 2.25 $\pm$ 0.3 \\
H Pfund 8 & 2667 $\pm$ 5 & 2.25 $\pm$ 0.3 \\
\hline
\end{tabular}
\end{table}

\begin{figure*}
\centering
\includegraphics[width=\hsize]{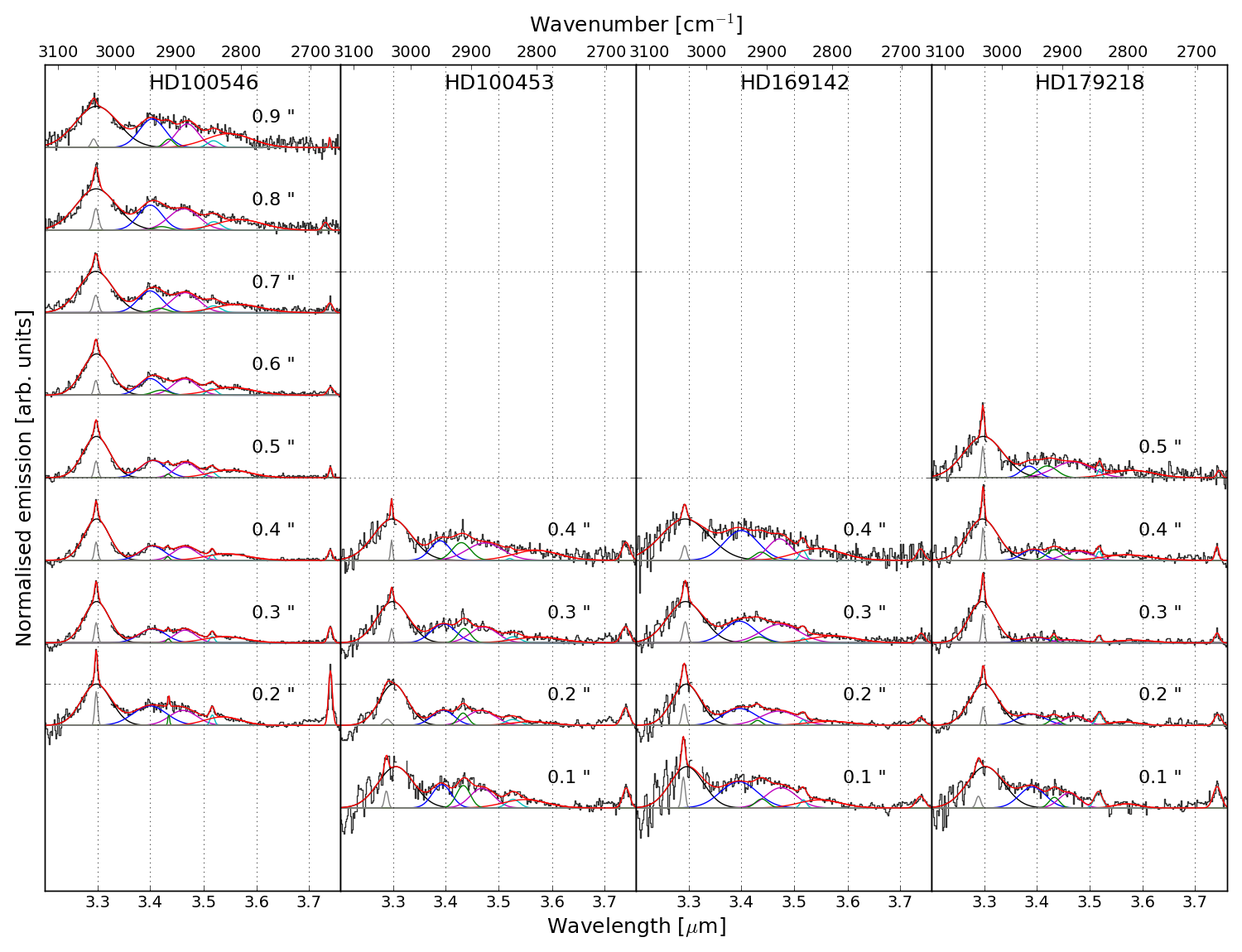}
\caption{NACO spectra averaged (in black) for different distances from the star in 0.1" step. Gaussian decomposition (in red) for each spectrum. Spectra are normalised to the continuum at 3.2~$\mu$m              }
\label{fig_globdec}
\end{figure*}

\begin{figure}
\centering
\includegraphics[width=\hsize]{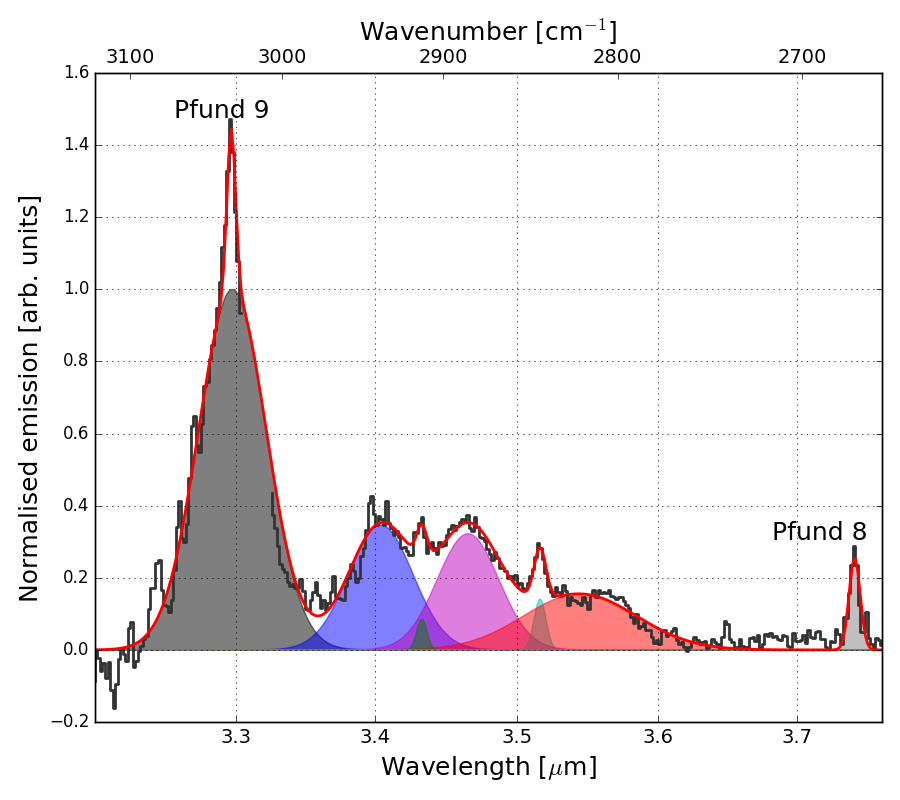}
\caption{NACO averaged spectrum (in black) of HD 100546 at 0.4" or 40 AU from the star. Gaussian decomposition is the red curve, hydrogen recombination lines are in grey. Individual contributor spectral signatures are identified at 3.3~$\mu$m (black), 3.4~$\mu$m (blue), 3.43~$\mu$m (green), 3.46~$\mu$m (magenta), 3.52~$\mu$m (cyan), 3.56~$\mu$m (red). The spectrum is normalised to the continuum at 3.2~$\mu$m. }
\label{fig_zdec}
\end{figure}

We optimize the parameters for each Gaussian using the Powell method. If the amplitude of the 3.3~$\mu$m band is three times lower than the mean noise of the underlying continuum, the spectrum is excluded. Then, we average the spectra according to their distance from the star, considering a step of 0.1" corresponding to the spatial resolution of the instrument. Fig.\ref{fig_globdec} presents the characteristic spectra according the distance to the central star. These are normalised to the continuum at 3.2 $\mu$m and thus show the evolution of the signatures relative to the continuum. 

The main observation is that all the disks exhibit similar spectral signatures: a 3.3 $\mu$m aromatic feature, and other features in the 3.4 - 3.6 $\mu$m range \ref{fig_globdec}. 

Figure \ref{fig_zdec} shows details for one characteristic spectrum. For the 3.3, 3.4, and 3.46~$\mu$m, the FWHM are similar whereas for the 3.43 and 3.52~$\mu$m bands, they are narrower. The FWHM of the 3.56~$\mu$m band is broader but the feature is poorly fitted. Hydrogen Pfund recombination lines are very close to their theoretical values, 3.297 and 3.741~$\mu$m with a narrow FWHM and confirm the good wavelength calibration of the data. The decomposition method gives similar results for the other spectra.

In the following, we will focus on similarities and variations of these signatures.

\subsection{Detection of the bands towards the disk \label{subsec_detection} }

Aromatic and aliphatic bands are detected in HD 100546 between 0.2 and 0.9" (20-100 AU) in the main outer disk (see Fig. \ref{fig_globdec}). 
In HD 100453, they are seen between 0.1 and 0.4" (10-40 AU) covering an area including both the gap that extends up to 20 AU and the external disk.
Likewise, in HD 169142, signatures are detected between 0.1 and 0.4" (10-45 AU) which includes part of the two gaps detected between 1 and 20 AU, and between 30 and 55 AU.
In HD 179218, bands are detected between 0.1" and 0.5" (30-120 AU). A gap exists out to 10 AU. 
Data do not allow us to probe the inner part of the disk. 
Interestingly, for most of the disks, the IR band carriers seem to be present both in the main disk regions and also in the gaps contrary to big grains (see references in the Appendix \ref{app_disks}).

\subsection{Intensity variations \label{subsec_signdis}}

\begin{figure}
\centering
\includegraphics[width=\hsize]{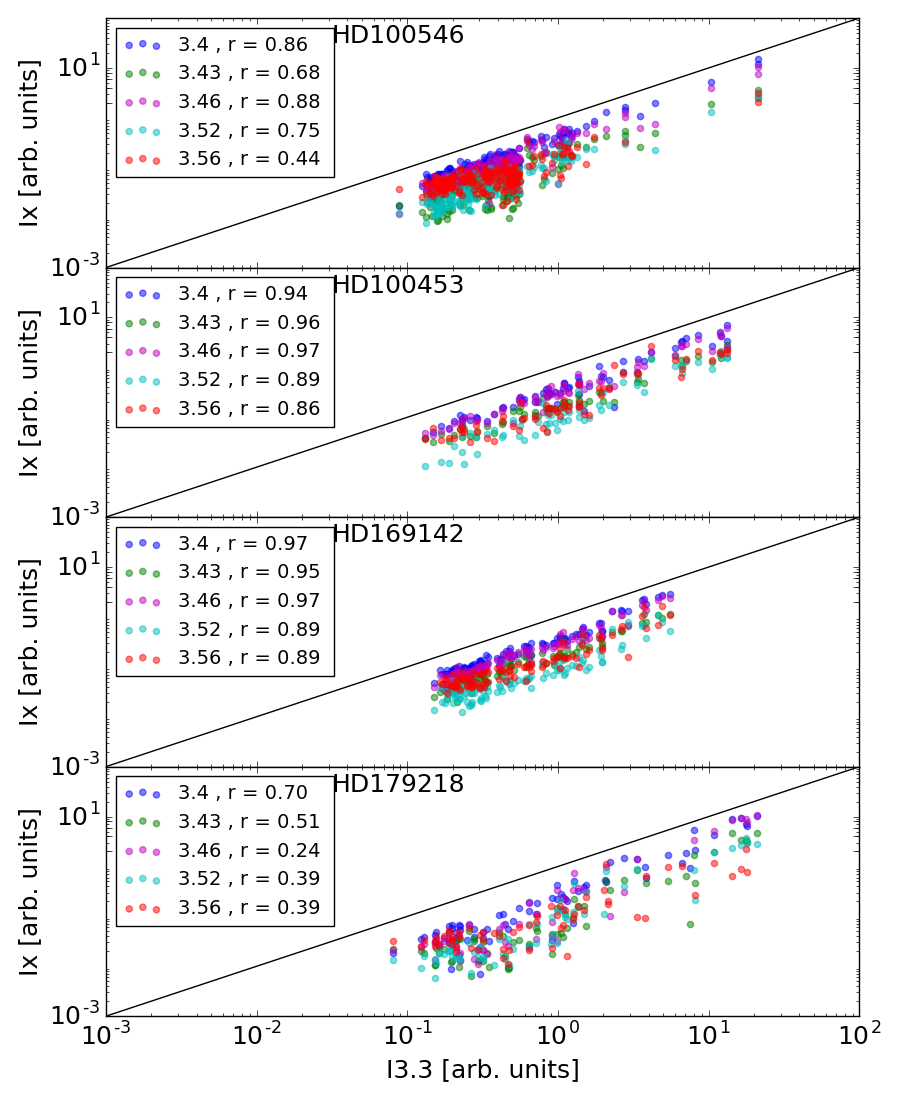}
\caption{Points are the intensity (amplitude $\times$ $\sigma$) of 3.4 (blue), 3.43 (green), 3.46 (magenta), 3.52 (cyan), 3.56 (red) $\mu$m band as function of the 3.3 $\mu$m band intensity. For each set, the Spearman coefficient r is calculated.}
\label{fig_correl}
\end{figure}

For each disk, Fig. \ref{fig_correl} shows the intensities of each  a-C:H emission band between 3.4 and 3.6 $\mu$m as a function of the intensity in the 3.3~$\mu$m aromatic band. These values come from the decomposition results of the individual pixel spectra that meet the selection criteria discussed previously.
For HD 100546, HD 100453, and HD 169142, the values are mostly linearly distributed. For HD 179218, the distribution is less clear. The Spearman rank correlation coefficients which describe the monotonic relationship between two variables show a good correlation between the 3.4 to 3.6~$\mu$m features and the 3.3~$\mu$m feature. 
The emission in aromatic and aliphatic bands in this wavelength range is very sensitive to the size of the particles. Also, finding such a spatial correlation suggests that the band carriers are both stochastically heated and have a similar size distribution.

\begin{figure*}
\centering
\includegraphics[width=\hsize]{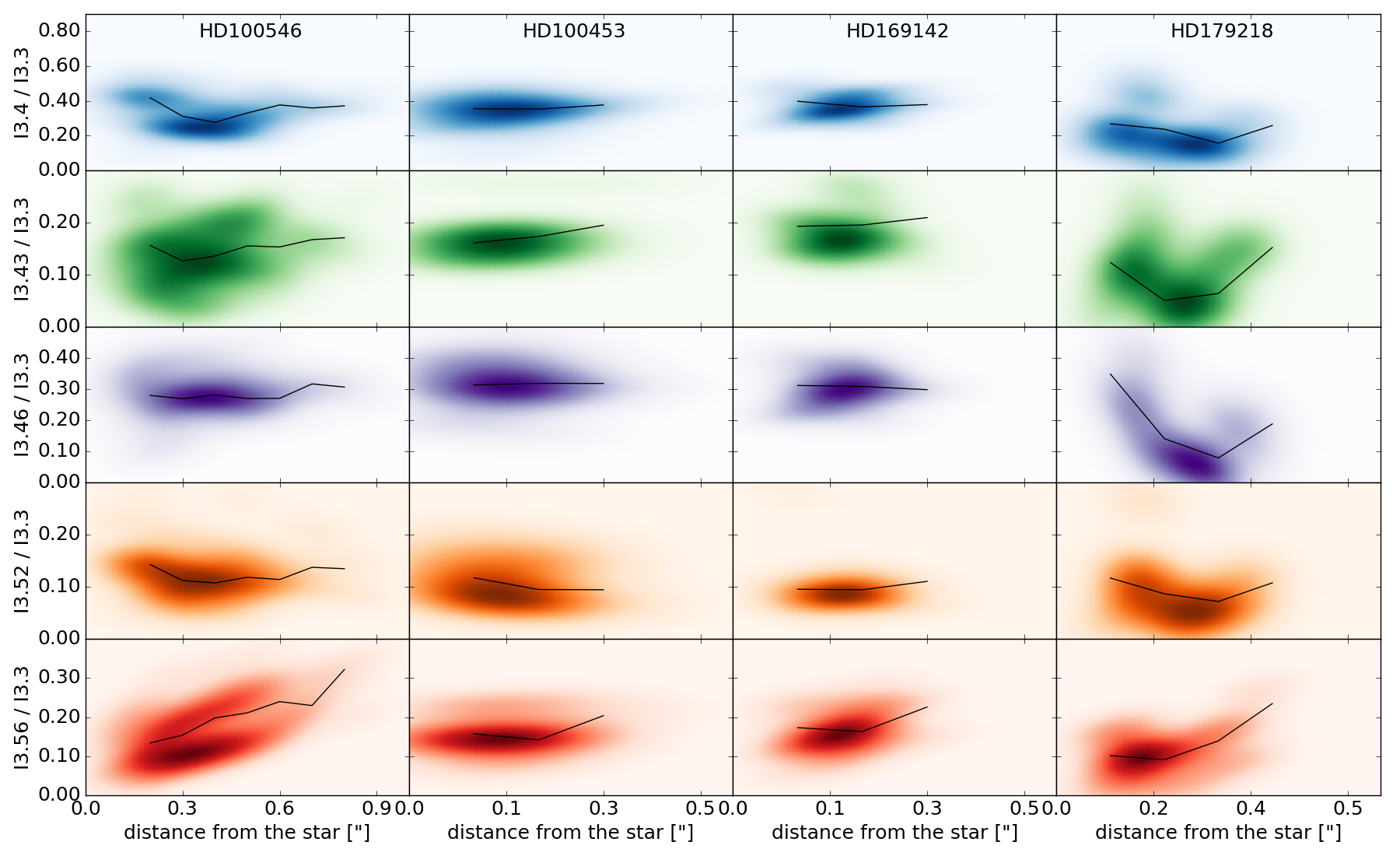}
\caption{Point densities are the band ratios (amplitude $\times$ $\sigma$) of 3.4 (blue), 3.43 (green), 3.46 (magenta), 3.52 (orange), 3.56 (red) to the 3.3~$\mu$m versus the distance from the star. Densities of points are plotted rather than points to make the variations more visible. The median value of the ratio is plotted as black solid line.}
\label{fig_aliaro}
\end{figure*}

Figure \ref{fig_aliaro} presents the distribution ratio values for the 3.4-3.6 $\mu$m signatures to the 3.3 $\mu$m band as a function of the distance from the star. 
For HD 100546, HD 100453 and HD 169142, the 3.4~/~3.3, 3.43 / 3.3, 3.46 / 3.3, and 3.52 / 3.3 ratios are of the same order and have a narrow dispersion without significant evolution as a function of the distance from the star, except for the ratio 3.43 / 3.3 for HD 100546 which has a larger dispersion. For HD 179218, the ratios are lower by \textasciitilde 1.4. As specified in Table \ref{tab_selobj}, HD 179218 is more luminous than the three other stars,  we expect carbonaceous materials to be differently processed due to the very different physical conditions. The aliphatic bonds, more fragile than the aromatic rings, are expected to be the first to break under UV processing. 
  
\subsection{The 3.3 $\mu$m band \label{subsec_aroband}}

In this section, we investigate the position and width of the 3.3~$\mu$m feature. 
Previous studies have pointed out that the 3.3~$\mu$m feature is actually composed of two sub-bands: one peaking at 3.28~$\mu$m and another at 3.30~$\mu$m  \citep{sadjadi_origin_2017,candian_spatial_2012,song_evolution_2003,tokunaga_high-resolution_1991,hammonds_variations_2015,kwok_mixed_2011}. 
In particular, \citet{tokunaga_high-resolution_1991} classified the 3.3~$\mu$m band in two types. Type 1 peaking at 3.289~$\mu$m with width of 0.042~$\mu$m and observed in numerous objects such planetary nebulae and HII regions, a type 2, peaking at 3.296~$\mu$m with narrower width of 0.020~$\mu$m which are observed in pre-main sequence and evolved stars. \cite{tokunaga_high-resolution_1991} noted that laboratory data did not allow a particular assignment to the features and proposed that Type 2 are best fitted by amorphous aromatic materials and Type 1 by heated PAHs. As suggested in Fig. \ref{fig_aliaro}, the carriers of the bands are likely due to a single material. 
A series of studies made a distinction between Type 1 and Type 2 concerning the 3.3~$\mu$m band and investigated its origin in the Red Rectangle in emission \citep{song_evolution_2003,candian_spatial_2012} and along a diffuse line of sight toward the Galactic centre \citep{chiar_structure_2013}. While \citet{candian_spatial_2012} proposed that the 3.28~$\mu$m feature comes from the structural organisation of the PAHs, \citet{chiar_structure_2013}  suggested an aliphatic origin to the 3.28~$\mu$m band due the stretching mode of olefinic C-H bonds in amorphous hydrocarbons.
Alternatively, \citet{sadjadi_origin_2017}  performed quantum mechanical calculations on a aromatic/aliphatic/olefinic compounds and proposed to assign the 3.28~$\mu$m feature to aromatics and  the 3.30~$\mu$m to olefinics.

The averaged band centre and width of the 3.3~$\mu$m band and variations with distance from the star are given in Table \ref{tab_arovar}. The band centre is in agreement with the laboratory experiments values reported in Table \ref{tab_versat}. For the all disks, the width, between 0.055 and 0.089~$\mu$m is globally higher than the highest width of 0.057~$\mu$m measured in the laboratory even if uncertainties could explain this difference in some cases.
The averaged band centre and width of the other features and their variation ranges are given in Table \ref{tab_alivar}.

\begin{figure*}
\resizebox{\hsize}{!}{\includegraphics[angle=0]{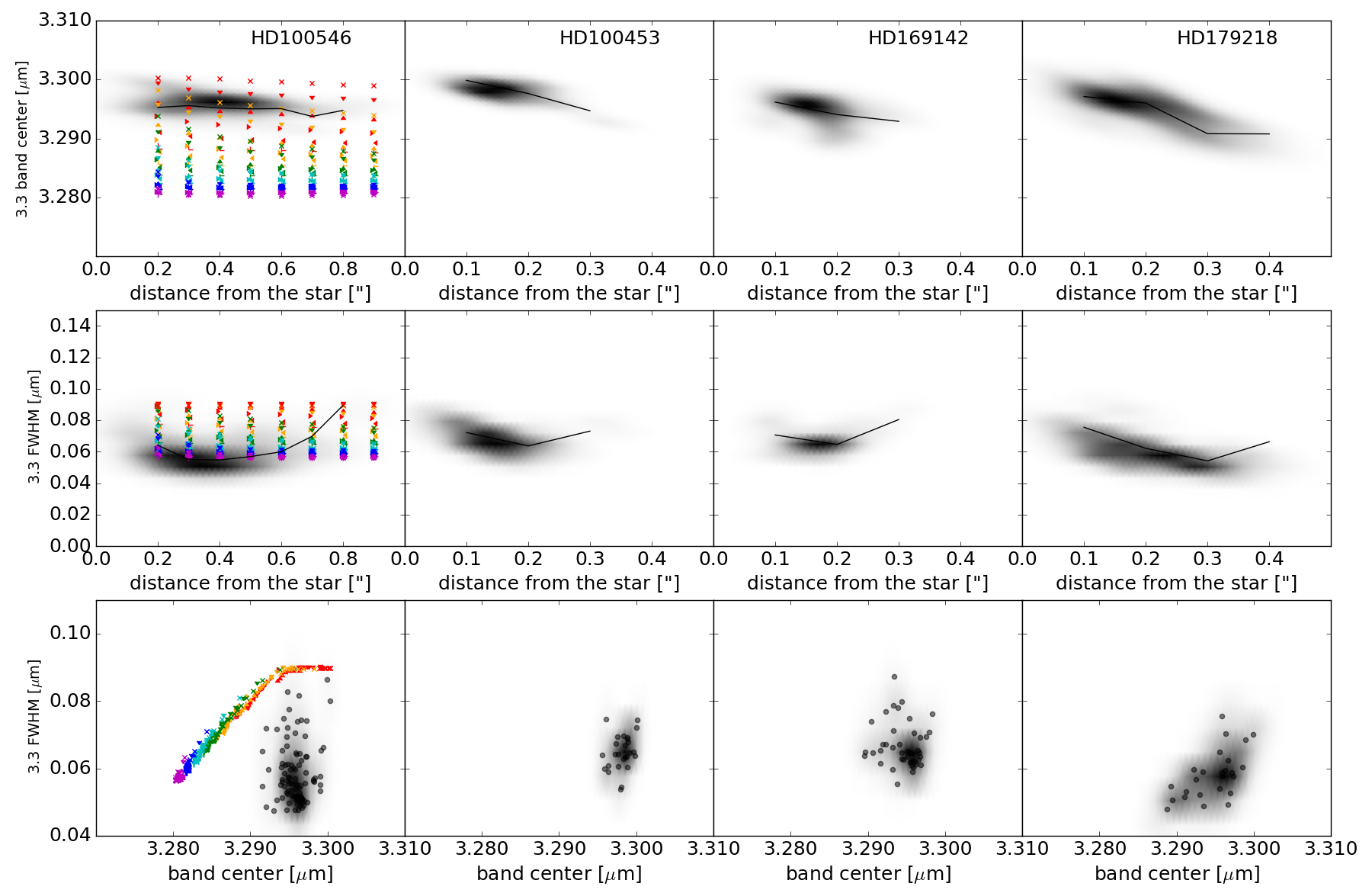}}  
\caption{In top panels, 3.3 $\mu$m band centre as a function of the distance from the star. In middle panels, 3.3 $\mu$m FWHM as a function of the distance from the star. In bottom panels, 3.3 $\mu$m FWHM over the band centre. For HD 100546, left panels, values are reported from THEMIS varying band gap from 0.1 eV (magenta) to 0.6 eV (red, as rainbow colours). Top two set of panel: Size increases bottom to top (for the same colour). Bottom panels: size increases from right to left.}
\label{fig_aroband}
\end{figure*}

In Fig. \ref{fig_aroband}, the parameter values of the band centre and FWHM coming from the decomposition of pixel spectra are plotted as a function of the distance from the star. In the bottom panels, the FWHM is plotted versus to the band centre. The variations are reported in Table \ref{tab_arovar}.
In HD 100453 and HD 179218, the average band centre decreases with distance from 3.30~$\mu$m to 3.29~$\mu$m. In HD 100546 and HD 169142, the band centre remains constant at 3.295~$\mu$m moving away from the star. 
In all the disks, the FWHM decreases then increases with distance from the star. We also note that in HD 179218, the FWHM increases for band centres at longer wavelength. This suggests that close to 3.28~$\mu$m, the carrier of the feature would be more aromatic and around 3.30 or 3.32~$\mu$m, it would be more olefinic/aliphatic \citep{sadjadi_origin_2017,chiar_structure_2013}. In this case, a narrow width would mean that one component dominates when a broader one would represent the presence of both components. This aspect will be explored in more detail in Sect. \ref{sec_mod}. 
For HD 100453 and HD 179218, the 3.3~$\mu$m band centre seems to shift from higher to lower values with distance from the central star. A priori, this is unexpected given the known behaviour of carbonaceous materials irradiated by UV photons. Close from the star, we expect that amorphous aliphatic materials would be aromatised by the more intense UV photon flux
\citep[e.g.,][]{jones_carbon_1990,dartois_diffuse_2004,dartois_organic_2004,pino_6.2_2008,mennella_activation_2008,godard_ion_2011}. Variations in the centre and width of the band suggest chemical processing (e.g. increasing contribution of olefinic/aliphatic bridges with increasing distance to the star) which are complex and difficult to interpret, requiring further study to allow definitive conclusions.
The 3.3~$\mu$m band evolution could thus be used as a diagnostic probe of the carbonaceous dust composition. 

\begin{table*}
\caption{3.3$~\mu$m band characteristics and variations as a function of the distance from the star (d).$^a$: too few data to calculate uncertainty. }             
\label{tab_arovar}      
\centering        
\begin{tabular}{c c c c c}     
  \hline \hline 
 & \multicolumn{2}{c}{HD 100546} &  \multicolumn{2}{c}{HD 100453} \\
d ["] &  $\lambda_{0}$ [$\mu$m] & FWHM [$\mu$m] & $\lambda_{0}$ [$\mu$m] & FWHM [$\mu$m] \\
\hline
0.1  & - - & - - & 3.300 $\pm$ 0.002 & 0.072 $\pm$ 0.008  \\
0.2  & 3.295 $\pm$ 0.004 & 0.065 $\pm$ 0.011 & 3.298 $\pm$ 0.004 & 0.064 $\pm$ 0.006  \\
0.3  & 3.296 $\pm$ 0.003 & 0.055 $\pm$ 0.005 & 3.294 $\pm$ 0.003 & 0.073 $\pm$ 0.005  \\
0.4  & 3.295 $\pm$ 0.003 & 0.055 $\pm$ 0.007 & 3.293 $\pm$ 0.001 & 0.089 $\pm$ 0.000$^a$  \\
0.5  & 3.295 $\pm$ 0.002 & 0.057 $\pm$ 0.008 & - - & - -  \\
0.6  & 3.295 $\pm$ 0.002 & 0.060 $\pm$ 0.012 & - - & - -  \\
0.7  & 3.294 $\pm$ 0.002 & 0.070 $\pm$ 0.010 & - - & - -  \\
0.8  & 3.295 $\pm$ 0.002 & 0.089 $\pm$ 0.013 & - - & - -  \\
0.9  & 3.295 $\pm$ 0.004 & 0.089 $\pm$ 0.002 & - - & - -  \\
\hline          
 &  \multicolumn{2}{c}{HD 169142} & \multicolumn{2}{c}{HD 179218} \\
d ["] & $\lambda_{0}$ [$\mu$m] & FWHM [$\mu$m] &  $\lambda_{0}$ [$\mu$m] & FWHM [$\mu$m] \\
\hline
0.1   & 3.296 $\pm$ 0.003 & 0.071 $\pm$ 0.010 & 3.297 $\pm$ 0.002 & 0.076 $\pm$ 0.012 \\
0.2   & 3.294 $\pm$ 0.003 & 0.065 $\pm$ 0.004 & 3.296 $\pm$ 0.003 & 0.062 $\pm$ 0.012 \\
0.3   & 3.293 $\pm$ 0.003 & 0.080 $\pm$ 0.009 & 3.291 $\pm$ 0.003 & 0.054 $\pm$ 0.003 \\
0.4  & 3.294 $\pm$ 0.002 & 0.089 $\pm$ 0.000$^a$ & 3.291 $\pm$ 0.002 &  0.066 $\pm$ 0.007 \\
0.5  & - - & - - & 3.290 $\pm$ 0.000$^a$ & 0.073 $\pm$ 0.000$^a$ \\
\hline             
\end{tabular}          
\centering        
\end{table*}

\subsection{The 3.4 to 3.3 $\mu$m band ratio\label{subsec_34aro}}

The top panel in Fig. \ref{fig_aliaro} shows values of the ratio between the 3.4 to 3.3 $\mu$m bands as a function of the distance from the star. Values of the averaged ratio are reported in Table \ref{tab_34aro}. The ratio does not vary significantly with the distance while the radiation field intensity at the disk surface decreases by several orders of magnitude.
 
\begin{table*}
\caption{Band centres, widths and variation ranges for each disk (d).}             
\label{tab_alivar}      
\centering        
\begin{tabular}{c c c c c c c c }     
  \hline \hline 
 \multicolumn{2}{c}{HD 100546} &  \multicolumn{2}{c}{HD 100453} & \multicolumn{2}{c}{HD 169142} & \multicolumn{2}{c}{HD 179218} \\
 $\lambda_{0}$ [$\mu$m] & FWHM [$\mu$m] &  $\lambda_{0}$ [$\mu$m] & FWHM [$\mu$m] & $\lambda_{0}$ [$\mu$m] & FWHM [$\mu$m] & $\lambda_{0}$ [$\mu$m] & FWHM [$\mu$m]\\
   \hline
3.394 $\pm$ 0.005 & 0.052 $\pm$ 0.006 & 3.391 $\pm$ 0.002 & 0.051 $\pm$ 0.004 & 3.389 $\pm$ 0.001 & 0.054 $\pm$ 0.001 & 3.390 $\pm$ 0.002 & 0.053 $\pm$ 0.004\\
3.430 $\pm$ 0.006 & 0.033 $\pm$ 0.008 & 3.432 $\pm$ 0.001 & 0.031 $\pm$ 0.003 & 3.431 $\pm$ 0.004 & 0.035 $\pm$ 0.004 & 3.432 $\pm$ 0.004 & 0.031 $\pm$ 0.007\\
3.468 $\pm$ 0.003 & 0.044 $\pm$ 0.007 & 3.469 $\pm$ 0.002 & 0.053 $\pm$ 0.004 & 3.470 $\pm$ 0.003 & 0.052 $\pm$ 0.004 & 3.469 $\pm$ 0.005 & 0.046 $\pm$ 0.008\\
3.515 $\pm$ 0.003 & 0.035 $\pm$ 0.007 & 3.520 $\pm$ 0.007 & 0.036 $\pm$ 0.008 & 3.515 $\pm$ 0.005 & 0.039 $\pm$ 0.007 & 3.516 $\pm$ 0.003 & 0.031 $\pm$ 0.004\\
3.560 $\pm$ 0.008 & 0.075 $\pm$ 0.013 & 3.561 $\pm$ 0.009 & 0.085 $\pm$ 0.006 & 3.562 $\pm$ 0.009 & 0.084 $\pm$ 0.009 & 3.569 $\pm$ 0.006 & 0.061 $\pm$ 0.018\\
\hline                  
\end{tabular}
\end{table*}

\begin{table*}
\caption{Ratios of the 3.4 to 3.3$~\mu$m bands for each disk. Their dispersion are also given.}             
\label{tab_34aro}      
\centering        
\begin{tabular}{c c c c c}     
  \hline \hline 
& HD 100546 & HD 100453 & HD 169142 & HD 179218 \\
d ["] & \multicolumn{4}{c}{3.4~/~3.3} \\
\hline  
0.1  & - - & 0.354 $\pm$ 0.122 & 0.397 $\pm$ 0.107 & 0.268 $\pm$ 0.165  \\
0.2  & 0.418 $\pm$ 0.124 & 0.352 $\pm$ 0.068 & 0.365 $\pm$ 0.058 & 0.236 $\pm$ 0.121  \\
0.3  & 0.311 $\pm$ 0.095 & 0.376 $\pm$ 0.058 & 0.378 $\pm$ 0.060 & 0.156 $\pm$ 0.085  \\
0.4  & 0.276 $\pm$ 0.105 & 0.320 $\pm$ 0.029 & 0.418 $\pm$ 0.049 & 0.258 $\pm$ 0.053  \\
0.5  & 0.330 $\pm$ 0.076 & - - & - - & 0.289 $\pm$ 0.000  \\
0.6  & 0.376 $\pm$ 0.087 & - - & - - & - -  \\
0.7  & 0.359 $\pm$ 0.041 & - - & - - & - -  \\
0.8  & 0.371 $\pm$ 0.120 & - - & - - & - -  \\
0.9  & 0.402 $\pm$ 0.076 & - - & - - & - -  \\
\hline                  
\end{tabular}
\end{table*}

A previous study by \citet{pilleri_mixed_2015} focusing on spatially resolved spectra of the PDR NGC 7023 with AKARI and showed an increase in the 3.4~/~3.3 ratio when the FUV flux G$_{0}$ decreases. In order to make a comparison between our results with those of \citet{pilleri_mixed_2015}, we applied our decomposition method (see Sect. \ref{sec_reduc}) to their data to calculate the 3.4~/~3.3 ratio. 
We obtain higher ratios but they vary in the same way with G$_{0}$ as found by \cite{pilleri_mixed_2015} (Table \ref{tab_pil}).

\begin{table}
\caption{Ratios of the 3.4 to 3.3$~\mu$m bands for NGC 7023.}             
\label{tab_pil}      
\centering        
\begin{tabular}{c c c c}     
  \hline \hline 
Position & G$_0$ & \multicolumn{2}{c}{3.4~/~3.3} \\
 & & \citet{pilleri_mixed_2015} & This work \\
\hline  
P1 & 7000 & 0.028 & 0.181 \\
P2 & 2600 & 0.090 & 0.318 \\
P3 & 200 & 0.11 & 0.330 \\
P4 & 150 & 0.13 & 0.336 \\
\hline                  
\end{tabular}
\end{table}

\begin{figure}
\centering
\includegraphics[width=\hsize]{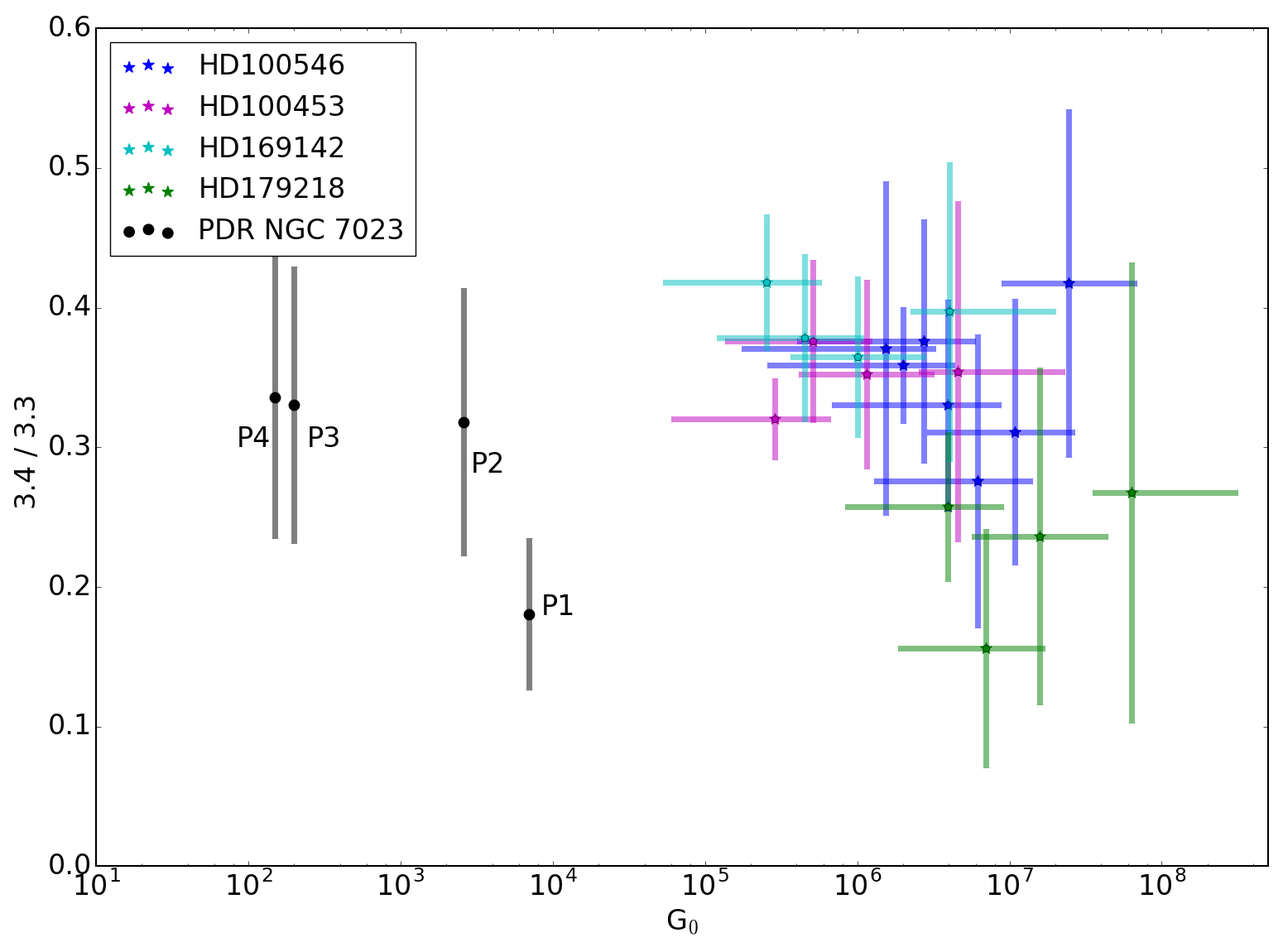}
\caption{3.4~/~3.3 ratio as a function of G$_0$. Black dot for band ratio in PDR NGC 7023 (based on \citet{pilleri_mixed_2015}. Stars for  band ratio for each disk and different distances to the central star. For the disk, the G$_0$ has been estimated from a black body associated to the central star parameters (temperature and luminosity). Error bars for the 3.4~/~3.3 ratio are the dispersion of ratio values for the considered distance. Error bars for G$_0$  correspond to the values for the considered distance inside the VLT/NACO beam.}
\label{fig_34arogz}
\end{figure}

Figure \ref{fig_34arogz} shows the 3.4~/~3.3 ratio versus G$_0$ for NGC 7023 and the four disks.
To calculate G$_0$ at different distances from the central star in disks, we have simply considered a blackbody with the effective temperature and stellar luminosity given in Table \ref{tab_selobj}.
The aliphatic-to-aromatic ratio does not appear to vary significantly between PDRs at the molecular edge and the disk surfaces. Assuming a constant gas density, with the orders of magnitude higher UV flux expected in our disk sample, with the Pilleri extrapolation, one would expect a very low 3.4/3.3 ratio, and a rapid dehydrogenation of the carriers following experiments. However, the emission in the 3-4~$\mu$m range depends not only on the UV flux but also on the (re-)hydrogenation rate, which in turn depends on the gas density. Thus, the observation of an aliphatic-to-aromatic ratio that varies little may suggests either a recent exposure of the carriers to the radiation field (by a continuous local vertical replenishment at the disk surface) before their destruction/conversion, and/or that the ratio of G$_0$/n$_H$ may be a better parameter to consider, and not the UV flux only. In disks, the gas density is expected to be much higher than in PDRs and must increase when we approach the star.

To go further in the understanding of the 3.4~/~3.3 ratio as a function of the local physical conditions, one can look at the intensity of the radiation in relation to the local gas density.
Based on the model of \citet{woitke_consistent_2016} considering a disk in thermal and hydrostatical equilibrium, the gas density at the disk upper surface at the radial optical depth $A_v=1$ can be estimated to be about $n_H=10^9$ and $10^7$ cm$^{-3}$ respectively at 10 and 100 AU from the central star (see their Fig. 4 for their reference model). Thus, considering that $G_0$ scales as the inverse squared distance to the illuminating star, one can estimate that G$_0$~/~n$_H$ does not vary to first order between 10 and 100 AU. 
This could explain a 3.4~/~3.3 ratio that does not vary much with distance from the star in the disks.
At the radial optical depth $A_v=1$ and considering $G_0^{50~AU}$ the FUV strength at 50 AU (see Table \ref{tab_selobj}), one can roughly estimate the G$_0$/ n$_H$ ratio as a function of the distance $d$ from the star by $\sim G_0^{50~AU} \times (\frac{50~AU}{d})^2 \times e^{(-A_v=1)}/ n_H(d)$.
We find G$_0$~/~n$_H\sim$0.04 for HD 100546, $\sim$0.002 and $\sim$0.004 for HD 100453 and HD 169142 and $\sim$0.15 for HD 179218 at 10 or 100 AU. 
The 3.4~/~3.3 ratios in HD 100546, HD 100453, and HD 169142, are comparable to those at the P2, P3, and P4 positions in the PDR NGC 7023.  
In HD 179218, the 3.4~/~3.3 ratio is between that found for the P1 and P2 positions in the cavity and the PDR edge. 

For NGC 7023,  estimates of G$_0$ can be found in \citet{pilleri_mixed_2015} and of n$_H$ (the total number of hydrogen nuclei) in \citet{kohler_hidden_2014}. 
This gives: (i) G$_0$/n$_H$= 10$^2$ in the cavity (ionised region - position P1); (ii) G$_0$/n$_H$= 0.26-2.6 at PDR's edge (position P2); (iii) G$_0$/n$_H$= 1-2 $\times$10$^{-2}$ in the PDR (A$_v\sim$1 - position P3); (iv) G$_0$/n$_H$= 10$^{-3}$-2.5$\times$10$^{-4}$ in the PDR ($A_v\sim$3 - position P4). The 3.4~/~3.3 ratio is low in the cavity (P1) and remains constant toward the PDR even if $G_0/n_H$ varies by several order of magnitudes.
Combining the PDR and disks results, we find that the 3.4~/~3.3 ratio is roughly equal to 0.3 - 0.4 for G$_0$/n$_H \le 10^{-2}$ while it decreases to $\sim$0.2 for G$_0$/n$_H \ge 10^{-1}$.

In conclusion, the 3.4~/~3.3 band ratios observed in disks are comparable to those observed in PDRs and interestingly do not depend on the FUV strength over density ratio beyond a certain threshold.
However, a more detailed study of the disk shape and irradiation conditions is needed to estimate more accurately the physical conditions (G$_0$, n$_H$) at the disk surfaces. Specific radiative transfer code for each disk with the appropriate structure (cavities, gaps) would be required.

\section{Modelling \label{sec_mod}}

The Heterogeneous dust Evolution Model for Interstellar Solids (THEMIS) \footnote{\url{https://www.ias.u-psud.fr/themis/}} is based on data derived from laboratory experiments and is described in detail elsewhere \citep{jones_variations_2012-1,jones_variations_2012-2,jones_variations_2012-3,jones_evolution_2013,kohler_hidden_2014,jones_global_2017}. 
THEMIS has been already used to analyse dense cloud observations \citep{kohler_dust_2015,jones_mantle_2016,ysard_mantle_2016}, dust in the Magellanic Clouds \citep{chastenet_modeling_2017} and other nearby galaxies within the Dustpedia project\footnote{http://dustpedia.com/}.
Given that we are here mostly interested in the emission coming from stochastically-heated nano-particles we briefly summarise the essential elements pertaining to these particles of THEMIS, which have radii from 0.4\,nm to several tens of nm, a power-law size distribution steeply declining in abundance with increasing radius. These nano-particles are primarily of amorphous, hydrogen-poor a-C composition and principally consist of poly-aromatic units linked together by olefinic/aliphatic bridges. These grains are rendered hydrogen-poor through the effects of photolytic dehydrogenation by the ambient interstellar UV radiation field in the diffuse ISM \citep{jones_variations_2012-1,jones_variations_2012-2,jones_variations_2012-3}. The composition, structure and optical properties of the THEMIS carbonaceous grains, containing aromatic, olefinic and aliphatic sub-components are calculated using the extended random covalent network (eRCN) model for H-rich a-C:H and a defective graphite (DG) model for H-poor grains \citep{jones_variations_2012-1,jones_variations_2012-2}.
The carbonaceous material properties are principally determined by their band gap with associated optical properties calibrated using the available laboratory data as described in detail in \cite{jones_variations_2012-1,jones_variations_2012-2}. 

For grains with radii $\leqslant$ 20 nm, the size of the aromatic clusters is necessarily limited. As derived in \cite{robertson_amorphous_1986}, the band gap depends on the size of these clusters. Thus, for a given composition, the smaller grains will have a larger gap than their bulk counterpart \citep{jones_variations_2012-2,jones_variations_2012-3}. This last point is taken into account in THEMIS with the definition of an effective band gap, which depends upon the particle size. Thus, and in order to avoid any ambiguity in the following, we adopt the bulk material band gap in our description, with the advisory that, in general, the effective band gap is larger than the bulk material band gap \citep{jones_variations_2012-3}. 

In THEMIS the spectroscopic assignments for the C$-$H and C$-$C vibrational modes were based on the available laboratory spectroscopic data \citep{jones_variations_2012-1}. The parametrized band characteristics are a compromise given the diversity encountered in laboratory data (see Fig. \ref{fig_versat}) and observed through the interstellar IR emission bands. Thus, the THEMIS IR spectral predictions are globally and qualitatively consistent with the observed laboratory and astronomical spectra.
In particular, THEMIS can self-consistently explain the $3.3\,\mu$m band and its associated side bands in the $3.4-3.6\,\mu$m region. However, the current model is not able to explain the details of the observed emission spectra, {\it i.e.}, the band positions are not always in agreement with the observed bands and, as attested by the data presented in Fig. \ref{fig_versat}, neither are any of the available laboratory data sets. Thus, it appears that conditions in the laboratory cannot yet be made to match all the environments for the IR carriers in space and that is why we have not yet achieved completely satisfactory spectral fits of the models to the data. 
Additionally, the THEMIS model predictions were made for dust in the diffuse ISM, whereas the best IR emission band spectra are generally for PDR regions with generally more intense and harsher radiation fields. 

In our exploration of the dust model parameter space, for a given disk as a function of distance, the THEMIS model grid contain almost 200 spectra which are calculated using DustEM\footnote{\url{https://www.ias.u-psud.fr/DUSTEM/}}, a numerical tool for dust emission and extinction \citep{compiegne_global_2011}.

Using THEMIS in its current form we have calculated a model grid for each disk and explore the nature of the THEMIS nano-particles as a function of three parameters: 
\begin{enumerate}
\item the distance from the star, which translates into the local radiation field expressed in terms of $G_0$,
\item the grain composition, in terms of the material effective band gap, $E_{\rm g}$, which is explored over the range $E_{\rm g} = 0.0-2.5$\,eV, {\it i.e}, from aromatic-rich materials ($E_{\rm g} \sim 0$\,eV) to aliphatic-rich, a-C:H like materials ($E_{\rm g} > 2$\,eV), and
\item variations in the minimum size of the carbonaceous nano-grain size distribution ranging from a$_{min}$ = 0.4 to 1.0\,nm. For greater sizes, the distributuion follow that of \cite{jones_evolution_2013}.
\end{enumerate}

\begin{figure*}
\centering
\includegraphics[width=\hsize]{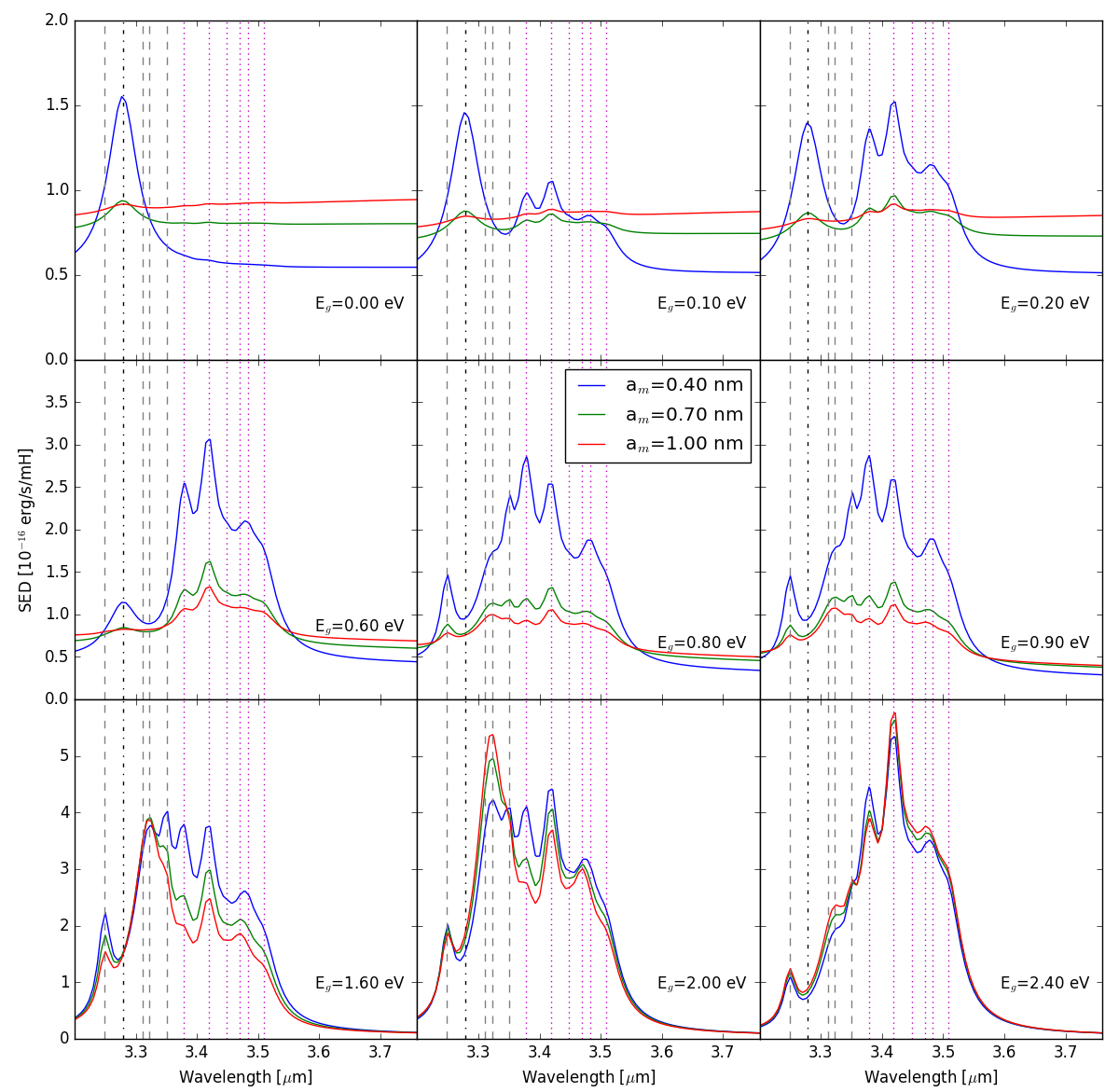}
\caption{ SED in the 3.2-3.8 $\mu$m range calculated from THEMIS for a radiation field intensity G$_0$ = 2.5$\times 10^7$ similar to that found in disks. Calculation made with DustEM. From top right to bottom left, band gap varies from 0.1 eV to 2.4 eV. In each subplot, SED is plotted for several minimal size of grains. }
         \label{fig_gridmod}
   \end{figure*}

Figure \ref{fig_gridmod} shows an example of the wide variability in spectral energy distribution (SED) allowed by THEMIS.
From the top left panel to the bottom right panel, the gap is increased,i.e., underlying carriers vary from aromatic materials (low gap) to aliphatics (high gap). Increasing the gap leads to a switchover of the aromatic signature (black dashdot) at 3.3 $\mu$m to olefincs/aliphatics signatures (grey dashed/magenta dotted) between 3.4 and 3.6 $\mu$m for higher gaps. On each panel color traces are SED with different minimal size of the smallest grains in the distribution size and illustrates the prevalence of the surface or bulk effects. The emission properties of the smallest grains vary rapidly with size and in particular the band/continuum ratio. The choice of this range of minimum size allows to illustrate these variations.
For small gaps, with increasing minimal size, we note an increase in the continuum emission and a decrease in the band emission due to both the surface/bulk effect and the effective lower gap for larger grains which are less limited by the size of aromatic clusters \citep{jones_variations_2012-1}.
For large gaps, as size increase, the continuum emission remains weak because the aromatic units are too small to allow emission at these wavelengths. 

Thus, a first exploration is made through the grid to compare to the observations. We focus on HD 100546 which is the disk for which spectra cover the largest range of distances from the star. The radiation field was modelled by a black body with the temperature and luminosity of the central star, {\it i.e.} 10500 K and 32 L$_{\odot}$. The field intensity G$_0$ was calculated for step of 0.1" (star located at 103 pc). Thus, for each distance from the star, the SED is simulated for several sets of parameters of band gap $E_{\rm g}$ and minimal size a$_{min}$. Then, to compare to the observations, the decomposition method presented in section \ref{sec_reduc} is used ont the model dat to get parameters of Gaussian functions. Model spectra are normalised to the observational spectrum at 0.2". 

The results are presented in Fig. \ref{fig_cormod_a5e1}. Figs. \ref{fig_cormod_a4e1} and \ref{fig_cormod_a5e2} in Appendix \ref{app_themis-grid} show the comparison with other sets of parameters.
In the left panel, results of the model decomposition are plotted (the similar aspect is due to the normalisation). On right panel, the relative intensities in the bands are reported on the same plot as in Fig. \ref{fig_correl}. 
For Fig. \ref{fig_cormod_a5e1}, where a$_{min}$ = 0.5 nm and Eg = 0.1 eV, we see that the intensities obtained after normalisation match the observations both in terms of relative order of distribution between the signatures for bands at 3.4, 3.43, 3.46, and 3.52 $\mu$m and in terms of intensities (derived from the field intensity G$_0$).
When the a$_{min}$ varies from 0.5 to 0.4 nm (Fig. \ref{fig_cormod_a4e1}), we observe a global shift to higher intensities. 
When the effective gap increases from 0.1 to 0.2  eV  (with a$_{min}$ = 0.5 nm, Fig. \ref{fig_cormod_a5e2}), intensities of aliphatic bands increase with respect to the 3.3 $\mu$m band.
The intensity of the signature at 3.56 $\mu$m seems to be underestimated when we consider a$_{min}$ = 0.5 nm and Eg = 0.1 eV.
As firstly seen in Sect. \ref{subsec_dec}, the fit of the 3.56 $\mu$m feature does not match very well. The decomposition method and THEMIS assume that observed signatures come from a-C:H materials that contain only carbon and hydrogen atoms. Thus, we speculate the 3.56 $\mu$m feature could be explained by the chemical composition of the dust, such as inclusions of hetero-atoms of nitrogen or oxygen. Indeed, as reported by \citet{tallent_carbon-hydrogen_1956,goebel_identification_1981}, some carbonaceous materials with nitrogen or oxygen do have signatures around 3.56 $\mu$m. In the particular case of possible oxygen inclusion, some features expected in the mid-IR wavelength range may be searched for. With the inclusion of nitrogen in the aromatic network, this would be less evident as the infrared activity of potential new modes may be weak, but could influence and shift the position of some of the mid-IR aromatic modes \citep{socrates_infrared_2004}.

Finally, as shown in Fig. \ref{fig_aroband} (bottom panel) in Sect. \ref{subsec_aroband}, we explore the width variation according to the band centre shift of the 3.3 $\mu$m band.
In THEMIS, both the increases in E$_g$ (i.e. change in the grain composition) and a$_{min}$ drive the increase in the width and the shift of the band centre to longer wavelengths. This argues for multiple components for the 3.3 $\mu$m band, from aromatic composition close to 3.28 $\mu$m and hybrid aromatic-olefinic composition close to 3.30 $\mu$m.

For the whole spectrum, we notice that the fit is not perfect. As explained above, THEMIS was developed for the diffuse ISM and signatures in this wavelength range present a great variability. Nevertheless, if the decomposition method does not use all signatures of the model, it fits signatures related to the same kind of materials (aliphatics) which evolve in a same way, as indicated by the correlation in Fig. \ref{fig_correl}. 
This first exploration allows us some interpretations of the dust composition and size. Keeping in mind that uncertainties remain high both on: 
\begin{itemize}
\item radiation field intensity, due especially to the large spatial scale covered by one spectrum where G$_0$ varies quickly,
\item analysis is based on averaged spectra for each distance and does not consider local variations due to the structure,
\end{itemize}
these results, by confirming that dust in HD 100546 is composed mostly of aromatic-rich sub-nanometer grains, with aliphatic and olefinic components, indicate that THEMIS is useful to model dust in disks and set limits on its composition and size. Nevertheless, a more detailed study has to be done to fully and definitively understand the observed variations and their origin.

\begin{figure*}
\centering
\includegraphics[width=\hsize]{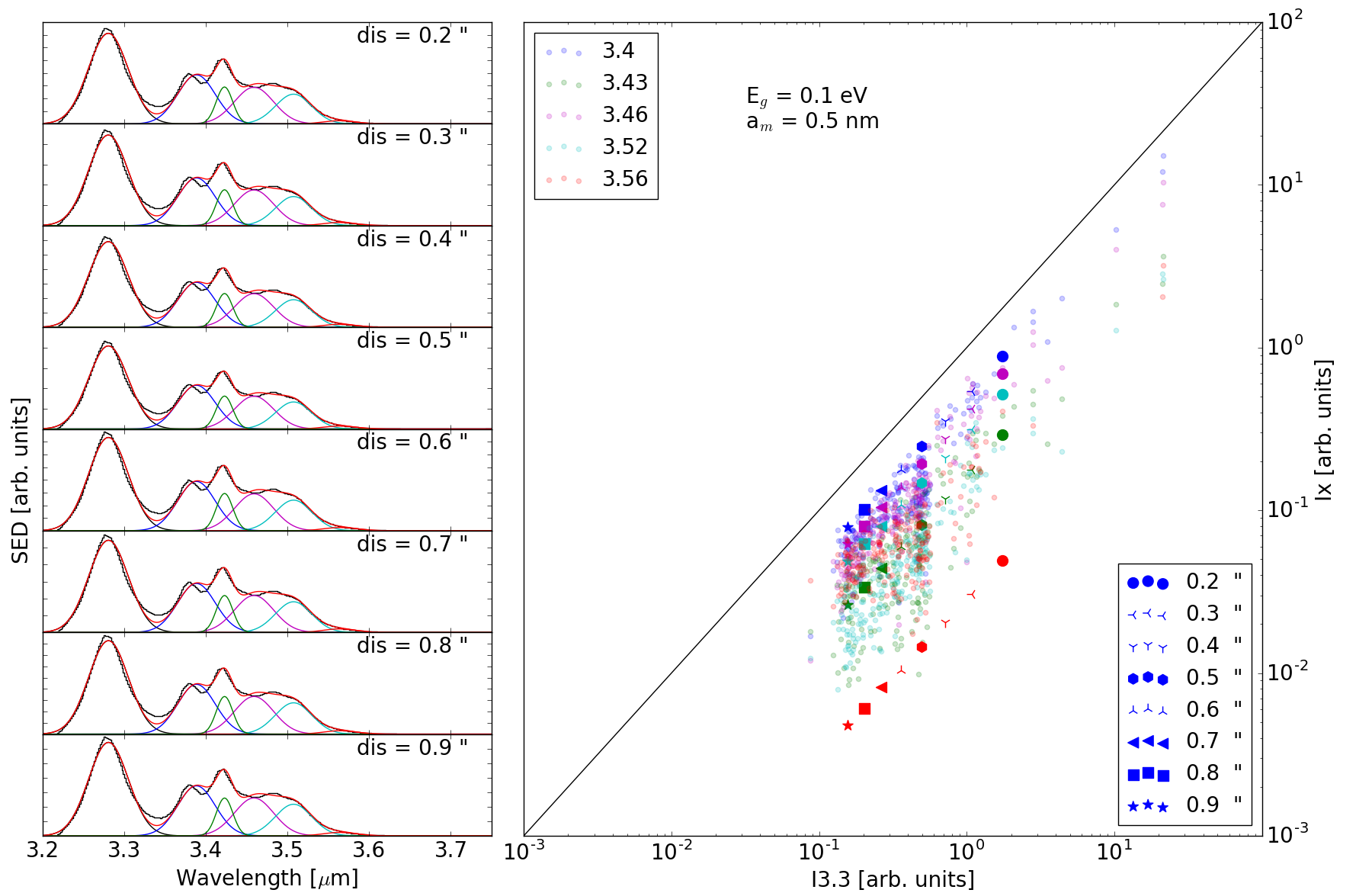}
\caption{In left panel, top down plots are the decomposition of the continuum subtracted spectra calculated with DustEM for a radial field intensity consistant with what is expected for HD 100546. All the spectra are normalised to the continuum at the distance at  0.2". In right panel, correlation between intensities in bands at 3.4, 3.43, 3.46, 3.52, 3.56 $\mu$m and the band at 3.3 $\mu$m underlined by those calculated for pixels of HD 100546 and previously shown in Fig. \ref{fig_correl}.}
\label{fig_cormod_a5e1}
\end{figure*}

\section{Conclusions \label{sec_conclusion}}

We present NACO/VLT spectroscopic observations in the 3 $\mu$m region of a small sample of HAeBe stars. These data allow us, for the first time, to spatially resolve the disk emission in the bands related to aromatic and aliphatic materials. Band assignments to aliphatic features is a complex question due to their high variability in terms of band positions and widths. 

\begin{enumerate}
\item This work confirms the presence of features related to aromatic and aliphatic materials in disks around Herbig stars. These features are observed in a wide spatial range of the disks (from about $\sim$10 to 100 au), even if the most inner parts of these disks remain inaccessible to our observations.
\item The various FUV radiation conditions and correlated intensities of aliphatic features with aromatic feature argue in favour of common nature of the carriers which are nano-grains stochastically heated. If there were distinct populations of carriers with significant different sizes, there would be no reason to have spatial correlated features.
\item No strong variations in band ratios (aliphatic / aromatic) according to the distance from the star are observed. This suggests features varying in a same way: same location, same intensity and argues for grains with mixed composition that is relatively stable. However, at a given distance, the dispersion of the values varies and depends on the data quality and the intrinsic decomposition variations but also on the physical dispersion due to the processes: aromatisation (due to UV irradiation), amorphisation, fragmentation and internal dynamics. An important point to consider is that the aliphatic bonds, more fragile than the aromatic rings, are expected to be the first to break under UV processing.  Thus, no strong variation in aliphatic -aromatic band ratios suggests that a continuous replenishment at the disk surface would be at work. This is supported by the fact that nano-grains of comparable size are observed throughout the disks, whereas one would expect the smallest grains to be destroyed in the most irradiated areas. Photo-induced and/or collisional fragmentation of larger grains that may be important at the disk surfaces could be a possible process to explain the replenishment of the nano-grains component. 
\item The 3.4/3.3 band ratios observed in disks are comparable to those observed in PDRs and appear to not only depend on the FUV strength but also on the gas density. Interestingly the 3.4~/~3.3 ratio does not depend on the FUV strength over density ratio beyond a certain threshold (of $G_0/n_H \le 0.01$). 
\item The band centre and FWHM variations in the 3.3 $\mu$m band suggest multiple components of the feature with aromatic/olefinic composition. But evolution depends on the disk and remains unclear.
\item This first comparison with THEMIS gives a reasonable but not perfect match to the observations: this is not surprising since THEMIS was developed for the diffuse ISM.
\item The relative order of IR features are well reproduced with THEMIS for nano-grain population with Eg = 0.1 eV and a$_{min}$ = 0.5nm.\\
\end{enumerate}

This work raises new questions about the nature of the observed signatures and their underlying carriers. 
In our follow-up work we will characterise the processing times of dust grains at the disk surface (photo-dissociation, recombination, dehydrogenation) to understand the process of their replenishment. 
Regarding the detection of the bands and sub-bands of carbon dust in its many forms, which can be weak on a strong continuum, the combination of high spatial and spectral resolution, and sensitivity is essential.
The forthcoming MATISSE/VLT and JWST observations will provide data that will allow us to further analyses and interpretations.

\begin{acknowledgements}
We thank our referee, A. Candian, for the care and quality of her comments and suggestions which greatly improved the clarity of this paper. This work was based on observations collected at the European Southern Observatory, Chile (ESO proposal number: 075.C-0624(A)) and was supported by P2IO LabEx (ANR-10-LABX-0038)»  in the framework "Investissements d'Avenir" (ANR-11-IDEX-0003-01) managed by the Agence Nationale de la Recherche (ANR, France),  Programme National "Physique et Chimie du Milieu Interstellaire" (PCMI) of CNRS/INSU with INC/INP co-funded by CEA and CNES. We also acknowledge Paolo Pilleri for sharing his observations of the PDR NGC7023 \citep{pilleri_mixed_2015}.
\end{acknowledgements}


\bibliographystyle{aa} 
\bibliography{ref.bib}


\begin{appendix}

\section{disks \label{app_disks}}

\subsection{HD 100546}
HD 100546 is one of the nearest very well studied Herbig Be stars \citep[d=109$\pm$4 pc ][]{gaia_collaboration_gaia_2018,lindegren_gaia_2016} . Based on HST and ground-based high-contrast images, an elliptical structure are detected and extended up to 350–380 AU \citep{augereau_hst/nicmos2_2001} and multipled-armed spiral patterns are identified \citep{grady_disk_2001,ardila_hubble_2007,boccaletti_multiple_2013}.
An inner dust disk extending from $\sim$0.2 AU to $\sim$1-4 AU was resolved using near-IR interferometry \citep{benisty_complex_2010,tatulli_constraining_2011,mulders_why_2013,panic_resolving_2014}.
The (pre-)transitional nature of HD 100546 was initially proposed by \cite{bouwman_origin_2003} from a SED analysis. The presence of a gap extending up to 10-15 AU has been confirmed from mid-IR interferometry \citep{liu_2003_resolved,panic_resolving_2014}, spectroscopy in the UV and the near-IR \citep{grady_disk_2007,brittain_tracing_2009,van_der_plas_evidence_2009}, and high-resolution polarimetric imaging in the optical and near-IR \citep{quanz_confirmation_2015,garufi_sphere_2016,follette_complex_2017}.

It is the first disk in which crystalline silicates \citep{hu_photometric_1989} and aromatic bands \citep{malfait_spectrum_1998} have been detected. 
HD100546 is one of the few HAeBe stars to show, at the same time, strong aromatic bands luminosity  and warm gas lines luminosities  \citep{meeus_observations_2012,meeus_digit_2013}.
The 8.6 and 11.3 $\mu$m aromatic features are spatially extended on a few 100 AU scale \citep{vanboekel_spatially_2004}, and the spatial distribution of the 3.3 $\mu$m emission shows a gap in the innermost region ($\sim$5–10 AU) and is extended up to at least 50 AU \citep{habart_spatially_2006}. The FHWM is about  $\sim12\pm 3$AU  \cite{geers_spatially_2007}. In \cite{habart_spatially_2006}, some additional features in the 3.4-3.5 $\mu$m region are also detected, that are at the time attributed to aliphatic C-H stretches in methyl or ethyl side-groups attached to PAHs \citep{joblin_spatial_1996,yang_carriers_2013,yang_c-h_2016} or nano-diamonds. 

\subsection{HD 100453}
The Herbig Ae star HD 100453A 
 \citep[d=104$\pm$3 pc,$M=1.7M_{sol}$][]{gaia_collaboration_gaia_2018,dominik_understanding_2003}, is less brighter and studied than HD 100546.  Whose proto-planetary disk was recently revealed to host a gap \citep{khalafinejad_large_2016} and a global two-armed spiral structure in SPHERE/VLT image consistent with a companion-driven origin \citep{wagner_status_2017}[and reference therein].
The primary A-star hosts an M-dwarf companion with a mass of $\sim0.2M_{\dotfill}$ and an angular separation of 1.05", corresponding to a projected physical separation of $\sim$108 AU if the orbit is seen close to face-on \citep{chen_vlt/naco_2006,collins_hd_2009}. 
The spatially resolved MIR Q-band image obtained with Gemini north/MICHELLE indicates an outer gap edge at $\sim$20 AU and the disk is extended up to $\sim$200 AU \citep{khalafinejad_large_2016}. 
From Gemini Planet Imager (GPI) polarized intensity (PI), SPHERE data and the IR SED fitting, \citet{long_shadow_2017} suggests that the circumstellar disk of HD 100453 appears to contain an inner disk which SED fitting suggests extends from 0.13 - 1 AU, followed by a large radial gap (1 - 18 AU), and an outer disk (18 - 39 AU).

Its Spitzer/IRS spectrum showing a very weak sign of silicate features at 10 and 20 $\mu$m indicates also the presence of a gap in the disk \citep[e.g.][]{marinas_high-resolution_2011,maaskant_identifying_2013}. 
The aromatic feature at 3.3, 6.2, 7.8, 8.6, 11.3 and 12.7 $\mu$m are detected \citep{meeus_iso_2001}, with secondary features observed at  5.7, 6.0, 10.6, 12.0, and 13.5 $\mu$m,
and the 6.8 and 7.2 $\mu$m aliphatic bands \citep{acke_spitzers_2010}.
HD 100453 also displays an emission feature at 16–19 $\mu$m, that could be attributed to the out-of-plane skeletal modes of large elongated PAH molecules \citep{peeters_polycyclic_2004,acke_spitzers_2010}.
The 3.3 $\mu$m aromatic feature is spatially extended with a FWHM equal to  $\sim$20 AU \citep{habart_spatially_2006,klarmann_interferometric_2017} presents NIR Interferometric observations showing extended flux that could be evidence for carbonaceous stochastically heated particles in the inner region of the proto-planetary disks around HD 100453.

\subsection{HD 169142} 

HD 169142 is a Herbig Ae star \citep[A5Ve;][]{keller_pah_2008} , well-studied at a distance of 117 $\pm$ 4 pc  \citep{grady_disk_2007,manoj_evolution_2006,gaia_collaboration_gaia_2018}. It shows an (almost face-on) disk with (i) multiple gaps (an inner  between $\sim$1 and $\sim$20 AU and a middle between $\sim$30 and $\sim$55 AU) (ii) dust rings at the edge of the gaps (at $\sim$20-30 AU and at 55-85 AU) and (iii) an outer disk \citep[e.g.][]{honda_observations_2013,quanz_gaps_2013,osorio_imaging_2014,momose_detailed_2015,fedele_alma_2017,ligi_system_2017,bertrang_magnetic_2017}.

\citet{seok_dust_2016,pohl_circumstellar_2017} also model dust population according to its location or its evolution. 
\citet{seok_dust_2016} performed a comprehensive modeling of its SED as well as the PAH emission features with porous dust and astronomical-PAHs and found that three dust populations and relatively small PAH molecules with an ionization fraction of 0.8 can explain the entire SED and the observed PAH features\footnote{Similar PAHs properties are derived by Seok \& Li for HD100453 and HD 179218, while for HD100546 relatively large PAHs with a small ionization fraction are found. The ionization fraction is the probability of finding a PAH molecule in a nonzero charge state \citep{li_modeling_2003} }.
In this disk, the aromatic feature at 3.3 $\mu$m is clearly detected  with the other aromatic bands 6.2, 7.8, 8.6, 11.3 and 12.7 $\mu$m \citep{meeus_iso_2001,sloan_mid-infrared_2005}, as well as, secondary features observed at  5.7, 6.0, 10.6, 12.0, and 13.5 $\mu$m, and the 6.8 and 7.2 $\mu$m aliphatic bands \citep{acke_spitzers_2010}.
The 3.3 $\mu$m aromatic feature appears to be observed in the inner cavity and is spatially extended with a FWHM of 0.3’’ or $\sim$30 AU  \citep[and this paper]{habart_spatially_2006}.

\subsection{HD 179218}

HD 179218 is at a distance of $\sim$290 pc \citep{gaia_collaboration_gaia_2018} with a B9 spectral type and a strong luminosity of $\sim$180 $L_{\dotfill}$. It harbors a circumstellar disk revealed through its IR excess and that is well known for its silicates dominated IR spectrum \citep{bouwman_processing_2001,schutz_mid-ir_2005,van_boekel_10_2005}. Its IR spectra are dominated by crystalline forsterite and enstatite rather than small, amorphous silicate grains, HD 179218 is known to have a large percentage of crystalline dust. This indicates dust processing in the circumstellar disk of HD 179218, and the presence of cold enstatite at $\sim$10 AU implies that enstatite is mostly produced in the inner regions and is transported outward by radial mixing \citep{van_boekel_10_2005}. A double-ring-like emission at 10 $\mu$m has been spatially resolved 
HD 179218 is at a distance of $\sim$290 pc \citep[e.g.,][]{fedele_structure_2008}, peaking at $\sim$ 1 and 20 AU, respectively. A gap at $\sim$ 10 AU is reported based on the MIR interferometry with VLTI/MIDI 
HD 179218 is at a distance of $\sim$290 pc \citep{menu_structure_2015}. 

The aromatic feature at 3.3~$\mu$m is detected, as well as, the other aromatic bands at 6.2, 7.8 and 8.6 $\mu$m \citep{meeus_iso_2001,acke_iso_2004,acke_spitzers_2010}. A very recent study HD 179218 is at a distance of $\sim$290 pc  \citep{taha_spatial_2018} has had for the first time spatially resolved aromatic emission of the disk. The average FHWM of the 8.6 and 11.3 $\mu$m is of 0.232" or 67 AU and 0.280" or 81 AU. Based on spatial and spectroscopic considerations as well as on qualitative comparison with IRS 48 and HD 97048, they favor a scenario in which PAHs extend out to large radii across the flared disk surface and are at the same time predominantly in an ionized charge state due to the strong UV radiation field of the central star. 
The 3.3 $\mu$m aromatic feature is detected from 0.1'' until 0.5'' or 150 AU from the central star. In this study, we find that the FWHM is about 0.2'' or $\sim$60 AU. 
\cite{kluska_multi-instrument_2018} presents NIR Interferometric observations showing extended flux that could be evidence for carbonaceous stochastically heated particles in the inner region of the proto-planetary disks around HD 179218.

\section{Comparison THEMIS with observations \label{app_themis-grid}}

\begin{figure}[H]	
\centering
\includegraphics[width=\hsize]{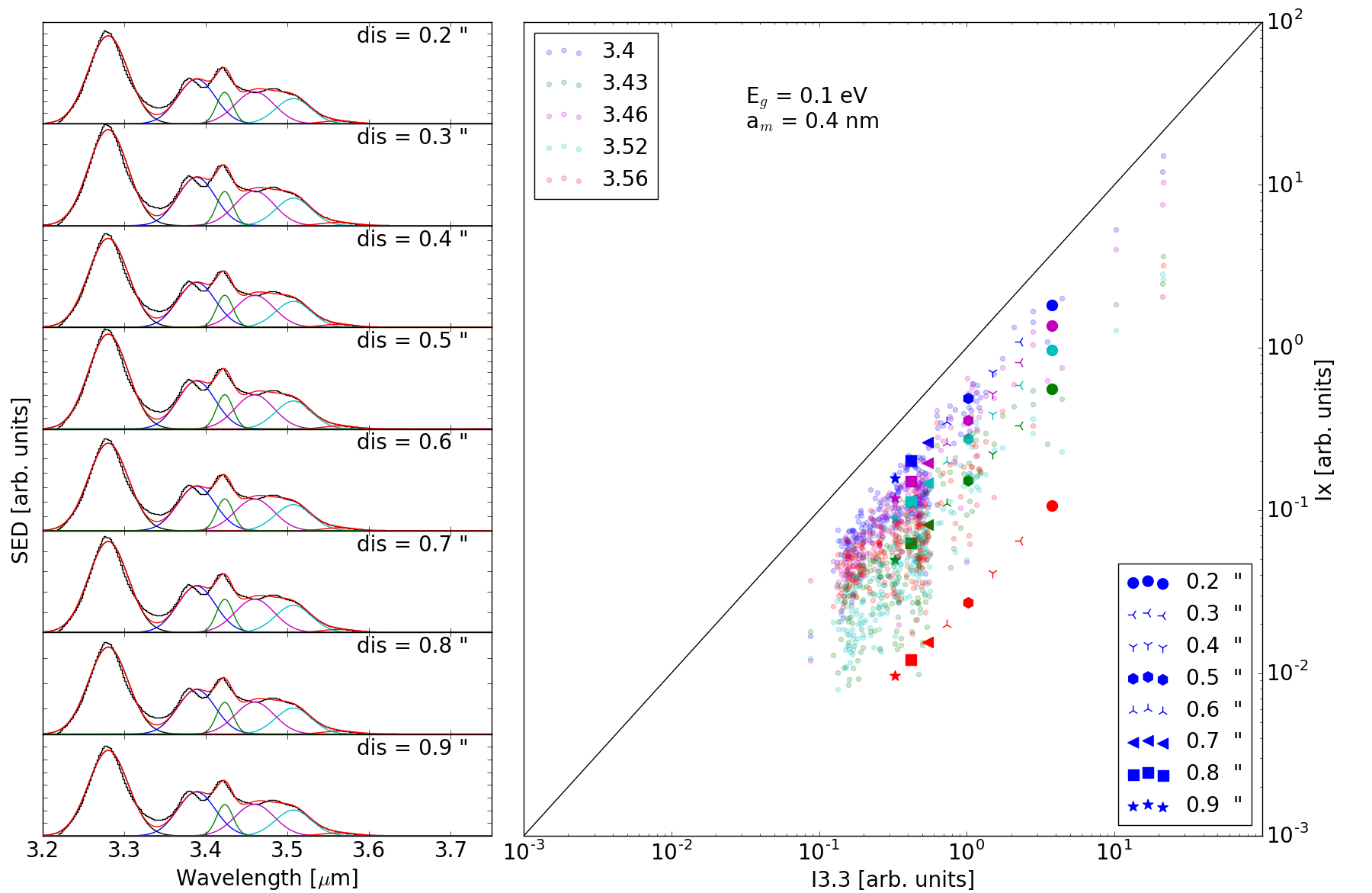}
\caption{This figure shows the correlation between intensities in bands 3.4, 3.43, 3.46, 3.52, 3.46 and the band at 3.3 $\mu$m underlined by those calculated for pixels of HD 100546 and previously shown in Fig. \ref{fig_correl}.}
\label{fig_cormod_a4e1}
\end{figure}

\begin{figure}[H]
\centering
\includegraphics[width=\hsize]{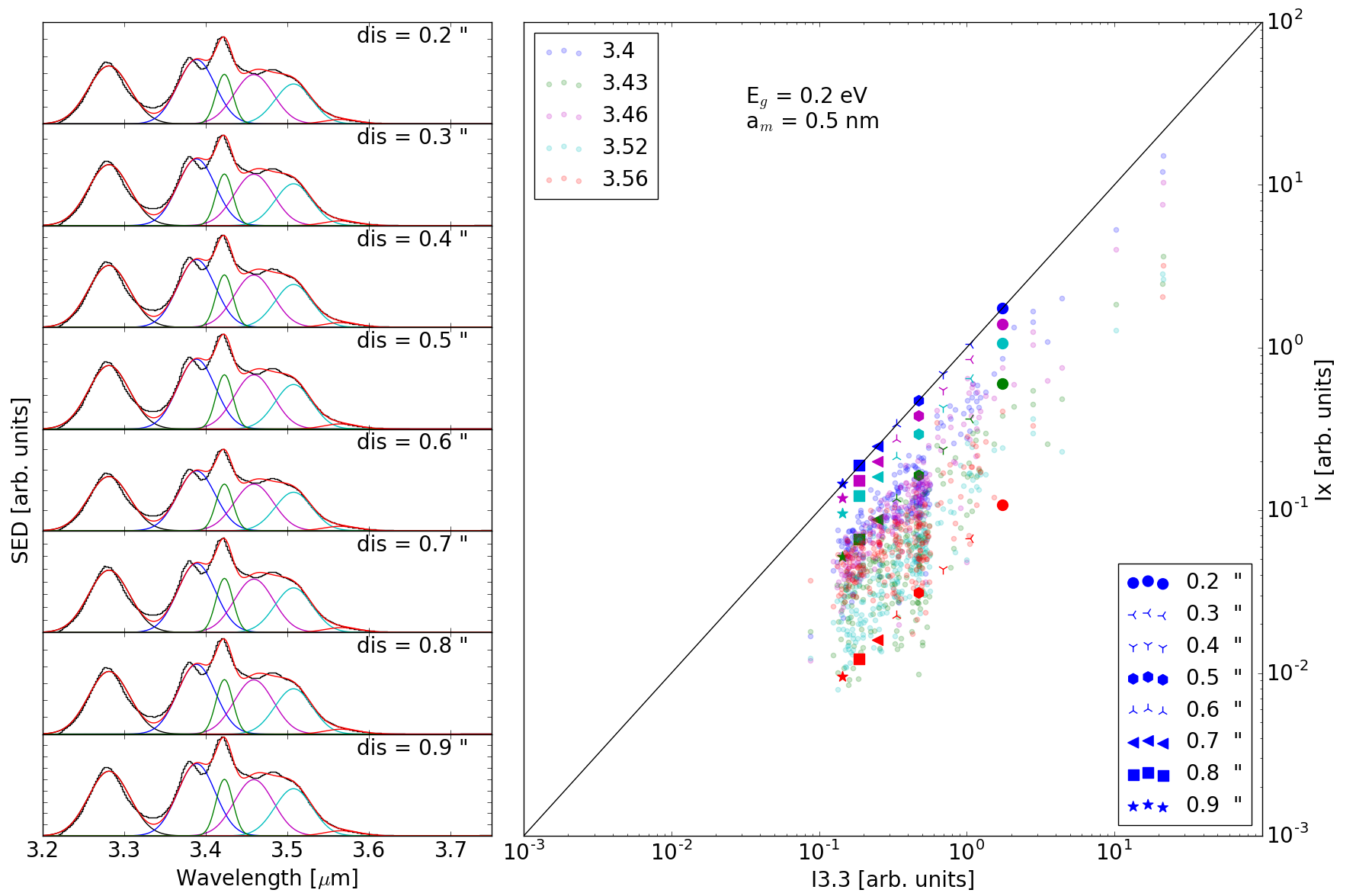}
\caption{This figure shows the correlation between intensities in bands 3.4, 3.43, 3.46, 3.52, 3.46 and the band at 3.3 $\mu$m underlined by those calculated for pixels of HD 100546 and previously shown in Fig. \ref{fig_correl}.}
\label{fig_cormod_a5e2}
\end{figure}

Figures \ref{fig_cormod_a4e1} and \ref{fig_cormod_a5e2} show correlations between the aliphatic bands and the aromatic one for two sets of parameters with THEMIS : (E$_g$=0.2 eV,a$_{min}$= 5 nm) and (E$_g$=0.1 eV,a$_{min}$= 4 nm), respectively. See Sec. \ref{sec_mod} for details

\end{appendix}

\end{document}